\documentclass[12pt]{article}
\usepackage{yfonts}
\usepackage{epsfig}
\usepackage{subfigure}
\usepackage{supertabular}
\usepackage{amsmath}
\usepackage{amssymb}
\usepackage{bbold}
\usepackage[sans]{dsfont}
\usepackage{graphicx}
\usepackage[dvips]{color}
\newcommand{\cleqn}{\setcounter{equation}{0}}
\newcommand{\newc}{\newcommand}
\newc{\om}{\omega}
\newc{\sig}{\sigma}
\newc{\beq}{\begin{equation}}
\newc{\eeq}{\end{equation}}
\newc{\beqn}{\begin{eqnarray}}
\newc{\eeqn}{\end{eqnarray}}
\newc{\bsym}{\boldsymbol}
\newc{\ol}{\overline}
\def\bvec#1{\raise1.5ex\hbox{$\rightarrow$}\mkern-16.5mu #1}

\def\ket#1{\vert\,#1>}
\def\m#1{\mathcal#1}

\begin{document}

\title{\hfill ~\\[-30mm]
       \hfill\mbox{\small UFIFT-HEP-07-2}\\[30mm]
       \textbf{The Flavor Group $\bsym{\Delta(3n^2)}$}}
\date{}
\author{\\Christoph Luhn,\footnote{E-mail: {\tt luhn@phys.ufl.edu}}~~
        Salah Nasri,\footnote{E-mail: {\tt snasri@phys.ufl.edu}}~~
        Pierre Ramond,\footnote{E-mail: {\tt ramond@phys.ufl.edu}}\\ \\
  \emph{\small{}Institute for Fundamental Theory, Department of Physics,}\\
  \emph{\small University of Florida, Gainesville, FL 32611, USA}}
\maketitle

\begin{abstract}
\noindent The large neutrino mixing angles have generated interest
in finite subgroups of $SU(3)$, as clues towards understanding the
flavor structure of the Standard Model.  In this work, we study the
mathematical structure of the simplest non-Abelian subgroup,
$\Delta(3n^2)$. Using simple mathematical techniques, we derive its
conjugacy classes, character table, build its irreducible
representations, their Kronecker products, and its invariants.
\end{abstract}

\vfill
\newpage
\section{Introduction}
At a time when the gauge structure of the Standard Model is well
understood, the origin of the triplication of chiral families
remains a mystery.  Several authors tried early on to explain chiral
family copies as manifestations of continuous family symmetries:
$SU(2)$\cite{GLM},  $SO(3)$\cite{WZ}, and $SU(3)$\cite{CF}.
Unfortunately, these theories which generically introduce more
parameters than need to be explained,  have met with at best partial
success.

Recent experimental clues have been added in the stew with the observation of neutrino oscillations where two large mixing angles have been measured.  This fact has led to a renewal of interest in the exploration of finite flavor groups,  where matrices with large angles appear naturally, as possible solutions to the flavor problem.

Many authors \cite{Pakvasa, Schmaltz, Ma, MNY, Ross, Kobayashi},
inspired by their favorite entries in the MNSP matrix, have
suggested specific finite groups, as holding the key to the flavor
problem. One of their guiding principles has been to focus on those
finite groups that are subgroups of a continuous flavor $SU(3)$.
These were listed long ago \cite{Miller}, and studied in some
details fifty years later \cite{Fairbairn:1964}.

However, there is no systematic approach to the study of  the finite groups that may explain flavor triplication. It is the purpose of this paper to provide a first step in that direction by presenting simple mathematical techniques for studying  the finite $SU(3)$ subgroups.

Three trivial finite $SU(3)$ subgroups are generated by $(3\times 3)$ diagonal matrices of unit determinant
\\
\\

$$\left(
  \begin{array}{ccc}
    1 & 0 & 0 \\
    0 & \alpha & 0\\
    0 & 0 & \overline\alpha \\
  \end{array}
\right) ,\qquad
\left(
  \begin{array}{ccc}
    \beta& 0 & 0 \\
    0 & 1 & 0\\
    0 & 0 & \overline\beta \\
  \end{array}
\right) , \qquad
\left(
  \begin{array}{ccc}
    \overline\gamma & 0 & 0 \\
    0 & \gamma & 0\\
    0 & 0 & 1 \\
  \end{array}
\right) ,$$
\\
\\
with $\alpha^m~=~\beta^n~=~\gamma^p~=~1\ ,$ generating the discrete group $\m{Z}_m\times \m{Z}_n \times \m{Z}_p$. A less
trivial finite group is one where the three entries of these diagonal matrices are permuted into one another. In that case, there are only two independent diagonal matrices
\\
\\
$$\left(
  \begin{array}{ccc}
    \beta& 0 & 0 \\
    0 & 1 & 0\\
    0 & 0 & \overline\beta \\
  \end{array}
\right) ,\qquad \left(
  \begin{array}{ccc}
    \overline\gamma & 0 & 0 \\
    0 & \gamma & 0\\
    0 & 0 & 1 \\
  \end{array}
\right) ,$$
\\ ~\\
where now $\beta^n~=~\gamma^n~=~1\ ,$ generating $ \m{Z}_n\times \m{Z}_n\ $.  When their entries are permuted  by
$\mathcal S_3$, the group of six permutations on the three diagonal entries, the finite group that ensues is called $\Delta(6n^2)$. In the simpler
case when the diagonal entries are permuted only by the three cyclic permutations, we obtain its finite subgroup $\Delta(3n^2)$.

In this paper, we use $\Delta(3n^2)$ to display general  techniques
for the study of finite groups. We derive its class structure,
character table, and irreducible representations. Using the
Kronecker products of its irreducible representations, we construct
the Clebsch-Gordan coefficients, and list its quadratic and cubic
invariants.\footnote{In the case of supersymmetry, renormalizability
allows only the quadratic and trilinear terms in the
superpotential.} Although most of these results can be found in the
literature \cite{Bovier:1980gc}, we hope that our exposition will
make the study of these finite groups accessible to particle
physicists.

\cleqn
\section{The Structure of $\bsym{\Delta(3n^2)}$}\label{sclass}
The group $\Delta(3n^2)$ is a non-Abelian finite subgroup of $SU(3)$ of order $3n^2$. It is isomorphic to the semidirect product of the cyclic group $\m{Z}_3$ with $(\m{Z}_n \times \m{Z}_n)$\cite{Bovier:1980gc},

$$\Delta(3n^2)~\sim~(\m{Z}_n \times \m{Z}_n)\rtimes \m Z_3\ .$$
Finite groups are most easily defined by their {\em presentation} in terms of their generators. $\Delta(3n^2)$ has three generators, $a$ which generates $\m Z_3$ with $a^3 = 1$,  $c$ and $d$ which generate $(\m Z_n\times\m Z_n)$, with $c^n = d^n =1$ and $cd = dc$.
$\m{Z}_3$ acts on $(\m{Z}_n \times \m{Z}_n)$  by similarity transformation (conjugation)\footnote{Different actions of $\m{Z}_3$ on $(\m{Z}_n
\times \m{Z}_n)$ yield semidirect products which are not $SU(3)$ subgroups.}

\beqn
a c a^{-1} ~=~  c^{-1} d^{-1} \label{def1a}\ ,\qquad
a d a^{-1} ~=~  c\ . \label{def1b}
\eeqn
Any of the $3n^2$ group elements $g \in \Delta(3n^2)$ can be written as a product of powers of $a$, $c$, and $d$

\beqn
g & = & a^\alpha c^\gamma d^\delta\ , \label{generalelt}
\eeqn
with $\alpha=0,1,2$ and $\gamma,\delta=0,1,...,n-1$.

\subsection{Conjugacy Classes}
The structure of finite groups derives from its conjugacy classes.
These are the set of group elements obtained from one another by
similarity transformations. Clearly, the identity element $e$  of
any finite group is always in a class by itself,\footnote{The prefix
before each class indicates the
  number of elements in the class.} $1C_1(e) = \{e\}$. From the presentation, we find
that

\beqn
cac^{-1} = ac^{-1}d, \qquad dad^{-1} = ac^{-1}d^{-2}\ ,
\label{action}
\eeqn
indicating that conjugacy does not alter the power of $a$. This enables us to study the class structure in terms of $a$. There are three types of classes;  those that contain:

\begin{itemize}
\item{{\bf{Elements without $\boldsymbol{a}$}}}

For fixed $\rho$ and $\sigma$,
where $\rho,\sig=0,1,...,n-1$ and $(\rho,\sig)\neq (0,0)$, the conjugacy class of the element
$c^{\rho}d^{\sigma}$ can be obtained by the action

\beqn g(c^{\rho}d^{\sigma})g^{-1}, \eeqn
leading to a class with at most three elements

\beq \{c^\rho d^\sig \,,\, c^{-\rho + \sig } d^{-\rho} \,,\,
c^{-\sig} d^{\rho -
  \sig} \}\ , \label{class1}
\eeq
obtained by a straighforward application of the presentation Eq.~(\ref{def1a}). The second element is obtained by conjugation with $a$

\beqn
a c^\rho d^\sig a^{-1} =(a c a^{-1})^\rho (a d a^{-1})^\sig  = (c^{-1} d^{-1})^\rho c^\sig =c^{-\rho + \sig} d^{-\rho}\ ,
\eeqn
and the third element by conjugation with $a^2$. It is easy to see that whenever the conditions

\beqn
3\rho~ = ~ \rho + \sigma~ =~ 0 ~\mathrm{mod}(n)\ ,\label{three}
\eeqn
are satisfied, the three elements in this class are the same, but this can occur only if $n$ is a multiple of three. Thus we consider two possibilities:

\begin{itemize}

\item[$(i)$]    $n\neq 3\,\mathbb{Z}$. The three elements in each class are all different. Although $\rho$ and $\sigma$ can take on $(n^2-1)$
possible values, the triples  $((\rho , \sig),(-\rho + \sig, -\rho),(-\sig,\rho - \sig))$
lead to the same class, leaving us with $(n^2-1)/3$ classes of three elements

\beqn
3\,C_1^{(\rho,\sig)} & =
& \{ c^\rho d^\sig \,,\, c^{-\rho + \sig } d^{-\rho} \,,\, c^{-\sig}
d^{\rho - \sig}\}\ .\label{class1aa}
\eeqn
\vskip .2cm
\item[$(ii)$] $n = 3\, \mathbb{Z}$.  Eq.~(\ref{three}) has two  solutions, giving  two classes with one element each. The remainder yields  $(n^2 -  3)/3$ three-element classes

\beqn
1\,C_1^{(\rho)} & = & \{ c^\rho d^{-\rho}
\}, ~~~~~~ \rho =\mbox{$\frac{n}{3} ,\frac{2n}{3}$} \ , \label{class1bb}\\
3\,C_1^{(\rho,\sig)} & = & \{ c^\rho d^\sig \,,\, c^{-\rho + \sig }
d^{-\rho} \,,\, c^{-\sig} d^{\rho -  \sig}\}, ~~~ (\rho,\sig) \neq
\left(\mbox{$\frac{n}{3},\frac{2n}{3}$}\right),
\left(\mbox{$\frac{2n}{3},\frac{n}{3}$}\right).~~~~~~~~\label{class1bc}
\eeqn
\end{itemize}
\vskip .3cm

\item {{\bf{Elements with $\boldsymbol{a}$}}}




Consider the element  $ac^{\rho}d^{\sig}$, with $\rho,\sig = 0,1,...,n-1$.
We discuss its conjugation  by  $g_\alpha =a^\alpha c^{\gamma}d^{\delta}$
separately for all three values of $\alpha=0,1,2$. The case
$\alpha=0$ yields, using Eq.~(\ref{action}),
\beqn
g_0 (a c^\rho d^\sig) g_0^{-1} &=& a
c^{\rho-\gamma-\delta}d^{\sig+\gamma-2\delta} \notag \\
&=& a c^{\rho +\sig - y_0-3x_0} d^{y_0},\label{ConjugCD0}
\eeqn
written in terms of the new parameters  $y_0 \equiv \sig+\gamma-2\delta$ and
$x_0 \equiv \delta$, with $x_0,y_0 = 0,1,...,n-1$. For $\alpha=1,2$ we get
recursively from Eq.~(\ref{ConjugCD0})
\beqn
g_\alpha (a c^\rho d^\sig) g_\alpha^{-1}
&=& a a c^{\rho +\sig - y_{\alpha-1}-3x_{\alpha-1}} d^{y_{\alpha-1}}  a^{-1} \notag\\
&=& a c^{\rho + \sig -y_\alpha -3x_\alpha} d^{y_\alpha},\label{ConjugCD12}
\eeqn
where $y_\alpha \equiv -\rho-\sig+y_{\alpha-1}+3x_{\alpha-1}$ and $x_\alpha
\equiv x_{\alpha-1} - y_\alpha$, with $x_\alpha,y_\alpha=0,1,...,n-1$.
The set of elements obtained by conjugation of  $ac^\rho d^\sig$ by $g_\alpha$
is therefore indepentent of $\alpha$. The corresponding
conjugacy class is given by the elements
\beq
a c^{\rho+\sig -y-3x} d^{y},\label{ConjugCD}
\eeq
with $x,y=0,1,...,n-1$.
This shows
that, for fixed $\rho$, $\sig$, and $y$, the exponent of $c$ can only change in steps of three mod\,$(n)$, leading us to two different cases:

\begin{itemize}

\item[$(i)$]  $n \neq 3\,\mathbb{Z}$.  All exponents of $c$ in Eq.~(\ref{ConjugCD}) can be
obtained just by varying $x$.  To see this, note that as $x$ takes its $n$ possible values, so does $3x~\mathrm{mod}\,(n)$; otherwise,
there would have to be an $x^\prime \neq x$, so that
$3(x^\prime -x) = 0~\mathrm{mod}\,(n)$, which is impossible for integer $x,x^\prime$.
We obtain only  one conjugacy class of $n^2$ elements

\beqn
n^2\,C_2 & = & \{ac^{x} d^{y}\, |\, x,y = 0,1,...,n-1 \}.\label{class2a}
\eeqn

\item[$(ii)$]  $n = 3\, \mathbb{Z}$.
In this case, only one third of all exponents can be obtained by varying $x$.
Thus we are led to three different conjugacy classes parameterized by $\tau=0,1,2$:

\beqn
\mbox{$\frac{n^2\!\!}{3}$}\, C_2^{(\tau)} & = & \{ac^{\tau-y-3x}
d^{y}\, |\, x = 0,1,..., \mbox{$\frac{n-3}{3}$};~y = 0,1,...,n-1\}\ .\label{class2b}
\eeqn
\end{itemize}
\vskip .3cm

\item {{\bf{Elements with $\boldsymbol{a^2}$}}}

This case is similar to the one above. The elements obtained from $a^2 c^\rho
d^{\sig}$  by the action of $g_\alpha = a^\alpha c^\gamma d^\delta$  are given
by



\beqn
g_0 (a^2 c^\rho d^\sig) g^{-1}_0
&=& a^2 c^{\rho -2\gamma+\delta} d^{\sig -\gamma - \delta} \notag\\
&=& a^2 c^{\rho +\sig - y_0 -3x_0} d^{y_0},
\eeqn
with $y_0 \equiv \sig -\gamma - \delta$ and $x_0 \equiv \gamma$ for
$\alpha=0$. In the case where $\alpha=1,2$ we find similar to Eq.~(\ref{ConjugCD12})
\beqn
g_\alpha (a^2 c^\rho d^\sig) g_\alpha^{-1}
&=& a^2 a c^{\rho +\sig - y_{\alpha-1}-3x_{\alpha-1}} d^{y_{\alpha-1}}  a^{-1} \notag\\
&=& a^2 c^{\rho + \sig -y_\alpha -3x_\alpha} d^{y_\alpha},
\eeqn
with $y_\alpha \equiv -\rho-\sig+y_{\alpha-1}+3x_{\alpha-1}$ and $x_\alpha
\equiv x_{\alpha-1} - y_\alpha$. The elements of the conjugacy class are thus
given by
\beq
a^2 c^{\rho +\sig -y-3x} d^y,\label{asquare}
\eeq
with $x,y=0,1,...,n-1$, which is to be compared with Eq.~(\ref{ConjugCD}).
The analysis proceeds as before:
\begin{itemize}

\item[$(i)$]  $n \neq  3\, \mathbb{Z}$. Use the same argument that led to
  Eq.~(\ref{class2a}), except for another factor of $a$, and obtain

\beqn
n^2 \,C_3 & = & \{a^2c^{x} d^{y}\, |\, x,y = 0,1,...,n-1 \}.\label{classc}
\eeqn
\item[$(ii)$] $n = 3\, \mathbb{Z}$.
 Here, from Eq.~(\ref{asquare}) the classes with elements containing  $a^2$
 are given by ($\tau = 0,1,2$)

\beqn
\mbox{$\frac{n^2\!\!}{3}$} \,C_3^{(\tau)} & = &
\{a^2c^{\tau-y-3x} d^{y}\, |\, x = 0,1,..., \mbox{$\frac{n-3}{3}$};~y = 0,1,...,n-1\}\ . \label{class3b}
\eeqn

\end{itemize}
\end{itemize}
\vskip .3cm

This completes the derivation of the class structure of the group $\Delta(3n^2)$. Our results can be summarized as:

\begin{itemize}

\item[($i$)]{$n \neq  3 \,\mathbb{Z}$.}~
Four types of classes

\beq 1\,C_1, ~~~ 3\,C_1^{(\rho,\sig)}, ~~~ n^2\,C_2, ~~~ n^2\,C_3, \eeq
adding up to $1+\frac{n^2-1}{3}+1+1$ distinct classes.

\item[($ii$)]{$n =  3 \, \mathbb{Z}$.}~
Five types of classes

\beq 1\,C_1, ~~~ 1\,C_1^{(\rho)}, ~~~
3\,C_1^{(\rho,\sig)}, ~~~ \mbox{$\frac{n^2\!\!}{3}$}\,C_2^{(\tau)},
~~~ \mbox{$\frac{n^2\!\!}{3}$}\,C_3^{(\tau)}, \eeq resulting in
$1+2+\frac{n^2-3}{3}+3+3$ different classes.

\end{itemize}

\cleqn
\section{Irreducible Representations}\label{sirreps}
Knowing the class structure, we construct the (unitary)
irreducible representations of $\Delta(3n^2)$. Thereafter, we can determine
the character table by taking the trace of the corresponding matrices.

\begin{itemize}
\item {{\bf{One-dimensional Representations}}}

In the one-dimensional representations, all generators commute with one another. Hence
 $a$ is solely constrained by $a^3 = 1$, 
giving

\beq a =1, \, \om, \, \om^2, ~~~~ \mathrm{with} ~~~ \om \equiv
e^{\frac{2\pi \mathrm{i}}{3}}\ .
\eeq
From Eq.~(\ref{def1a}),  we immediately see that

\beq
d=c\ , \qquad  c^3=1.
\eeq
This is compatible with $c^n=d^n=1$ only if  $c=1$ or $n$ is divisible by three. Hence the two cases:

\begin{itemize}

\item[$(i)$] $n \neq 3\, \mathbb{Z}$. $c$ and $d$ are necessarily one, and there are only three singlets
\beq
{\bf 1}_{r}: ~~~~~ a=\om^r,~~~ c=d=1,~~~~~ \mathrm{with} ~~ r=0,1,2\ .
\eeq

\item[$(ii)$]
$n=3\,\mathbb{Z}$.  $c=1,\om,\om^2$, leading to nine one-dimensional representations
\beq
{\bf 1}_{r,s}: ~~~~~ a=\om^r,~~~ c=d=\om^s,~~~~~ \mathrm{with} ~~ r,s=0,1,2\ .
\eeq
\end{itemize}

\item {{\bf{Three-dimensional Representations}}}

Using the method of induced representations, detailed in Appendix
\ref{inducedREP}, it is easy to show the existence of irreducible
three-dimensional representations. Since the generators $c$ and $d$ commute, they can be represented by diagonal $(3\times 3)$
 matrices with entries which are powers of the $n$-th root of unity $\eta \equiv e^{\frac{2\pi\mathrm{i}}{n}}$.
 We find for any two integers\footnote{excluding the values of $k$ and $l$ that lead to {\it reducible} representations, which happens
 if $k=l$ and  $3 k = 0~\mathrm{mod}(n)$.} $k, l=0,1,...,(n-1)$

\beq
a = \begin{pmatrix} 0 & 1 & 0  \\  0 & 0 & 1 \\ 1 & 0 & 0  \end{pmatrix},~~~~
c = \begin{pmatrix} \eta^l & 0 & 0  \\  0 & \eta^k & 0 \\ 0 & 0 & \eta^{-k-l}
\end{pmatrix},~~~~
d = \begin{pmatrix} \eta^{-k-l} & 0 & 0  \\  0 & \eta^l & 0 \\ 0 & 0 & \eta^k
\end{pmatrix}\ .\label{3drep}
\eeq
However, we must be careful not to count equivalent representations more than once. The easiest way to avoid overcounting is to note that the representations obtained by conjugation with $a$ and $a^2$ are equivalent.

\beq
c^\prime \equiv a c  a^{-1} = \begin{pmatrix} \eta^k & 0 & 0  \\  0 &
  \eta^{-k-l} & 0 \\ 0 & 0 &   \eta^l \end{pmatrix},~~~~
d^\prime \equiv a d  a^{-1} = \begin{pmatrix} \eta^l & 0 & 0  \\  0 & \eta^k
  & 0 \\ 0 & 0 &   \eta^{-k-l} \end{pmatrix}.\label{3dequivalentrep}
\eeq
Therefore the representation labeled by the pair $(k,l)$ is equivalent to that labeled
 $(-k-l,k)$. Conjugation by $a^2$ yields another equivalent labeling. The same representation is thus labeled in three different ways

\beq
\begin{pmatrix} k \\ l \end{pmatrix}, \begin{pmatrix} -k-l \\ k \end{pmatrix},
\begin{pmatrix} l \\ -k-l \end{pmatrix}.\label{3pairs}
\eeq
\newline

The  matrix

\beq
M ~\equiv~ \begin{pmatrix} -1 & -1 \\ 1 &0 \end{pmatrix}\ ,\qquad M^3~ =~ 1\ ,
\eeq
allows us to write these three pairs in the form

\beq
M^{p} \, \begin{pmatrix} k \\ l \end{pmatrix}, ~~~~\mathrm{with} ~~~ p=0,1,2.
\eeq
We have encountered a similar situation in labeling the classes $C_1^{(\rho,\sig)}$
by the pairs $(\rho,\sig)$ [see above Eq.~(\ref{class1aa})]. There, we  generated the three pairs of the triple by
$(\rho,\sig) M^p$, with $p=0,1,2$. When counting the number of classes with
three elements we introduced a factor of $1/3$ which takes care of the
labeling ambiguity of $C_1^{(\rho,\sig)}$.
To avoid the labeling ambiguity
of the three-dimensional irreducible {\it representations}, it is useful to
selects a standard representative of the three pairs in Eq.~(\ref{3pairs}). One possible choice is provided in Appendix~\ref{standard}.
Here we simply assume the existence of  a mapping $\widetilde{\phantom{wa}}$ that gives the standard representative:

\beq
\widetilde{\begin{pmatrix} k \\ l \end{pmatrix}} ~~ \longmapsto
~~\mathrm{either} ~~
\begin{pmatrix} k \\ l \end{pmatrix}, ~\mathrm{or}~~
\begin{pmatrix} -k-l \\ k \end{pmatrix},  ~\mathrm{or} ~~
\begin{pmatrix} l \\ -k-l \end{pmatrix},\label{mapping}
\eeq
depending on the values of $k$ and $l$. This leads us to the following inequivalent and irreducible three-dimensional representations:

\begin{itemize}
\item[($i$)]{$n \neq  3 \, \mathbb{Z}$.}

\beq
{\bf 3}_{\widetilde{(k,l)}}, ~~~~~ \mathrm{with} ~~ (k,l) \neq (0,0).
\eeq
Due to the mapping of Eq.~(\ref{mapping}) there are $\frac{n^2-1}{3}$ different three-dimensional representations of this type. Under
the implicit assumption that $(k,l)$ take only their standard values (see e.g. Appendix~\ref{standard}) we can omit the $\widetilde{\phantom{wa}}$ and
simply write ${\bf 3}_{(k,l)}$. Then the generators $a,c,d$ are given as in Eq.~(\ref{3drep}).

\item[($ii$)]{$n =  3 \, \mathbb{Z}$.}

\beq
{\bf 3}_{\widetilde{(k,l)}}, ~~~~~ \mathrm{with} ~~ (k,l) \neq (0,0),
(\mbox{$\frac{n}{3}$},\mbox{$\frac{n}{3}$}),
(\mbox{$\frac{2n}{3}$},\mbox{$\frac{2n}{3}$}).
\eeq
This gives $\frac{n^2-3}{3}$ three-dimensional representations. For notational ease, we will also write  ${\bf 3}_{(k,l)}$ instead of
${\bf 3}_{\widetilde{(k,l)}}$ in this case. In both cases, the complex conjugate representation is obtained by reversing $k$ and $l$ mod\,$(n)$.
\end{itemize}

There are no other irreducible representations. This can be easily seen from the formula which relates the order of the group to the sum of the squared dimensions of its irreducible representations

\beqn 3n^2 &=&
\sum_{\mathrm{irrep}~i} d_i^2 ,\label{counting}
\eeqn
where $d_i$
denotes the dimension of the irreducible representation $i$. Knowing all one- and three-dimensional representations we can check
Eq.~(\ref{counting}) to see if there are more irreducible representations. For $n
\neq 3\,\mathbb{Z}$ we have identified 3 one-dimensional and $\frac{n^2-1}{3}$ three-dimensional representations. Plugging this
into the right-hand side of Eq.~(\ref{counting}), we obtain $3 \cdot 1^2 + \frac{n^2-1}{3} \cdot 3^2 = 3n^2$.  Similarly we get for the
case where $n = 3\,\mathbb{Z}$ that $9 \cdot 1^2 + \frac{n^2-3}{3} \cdot 3^2 = 3n^2$.
\end{itemize}

We can now deduce the $\Delta(3n^2)$ character table by taking
traces over the relevant matrices, for both $n \neq 3\, \mathbb{Z}$
and $n = 3\, \mathbb{Z}$. The results are displayed in
Table~\ref{characterA}, using the classes and representations just
derived, including the restrictions on the parameters $(\rho,\sig)$
and $(k,l)$.

\begin{table}[htb]
\begin{center}
\subtable{ (a)~~~
\begin{tabular}{c||c|c|c|c} $\phantom{\Big|}n \neq 3 \, \mathbb{Z}$
& $1~C_1$ & $3~C_1^{(\rho,\sig)}$ & $n^2~C_2$ & $n^2~C_3$ \\ \hline
\hline $\phantom{\Big|}{\bf 1}_r$ & $1$ & $1$ & $\om^r$ & $\om^{-r}$
\\ \hline $\phantom{\Big|}{\bf 3}_{(k,l)}$ & $3$ & $\sum_{p}
\eta^{(\rho,\sig) M^p
  \mbox{\tiny$\begin{pmatrix} k \\ l \end{pmatrix}$}{\phantom{\big|}}}$ & $0$
& $0$
\end{tabular}}
\end{center}

\begin{center}
\subtable{ (b)~~~
\begin{tabular}{c||c|c|c|c|c}
$\phantom{\Big|}n = 3 \,\mathbb{Z}$ & $1~C_1$ & $1~C_1^{(\rho)}$ &
$3~C_1^{(\rho,\sig)}$ & $\mbox{$\frac{n^2\!\!}{3}$}~C_2^{(\tau)}$ &
$\mbox{$\frac{n^2\!\!}{3}$}~C_3^{(\tau)}$ \\ \hline \hline
$\phantom{\Big|}{\bf 1}_{r,s}$ & $1$ & $1$ & $\om^{(\rho + \sig)s}$
& $\om^{r+\tau s}$ & $\om^{-r+\tau s}$\\ \hline $\phantom{\Big|}{\bf
3}_{(k,l)}$ & $3$ & $\sum_{p} \eta^{(\rho,-\rho) M^p
  \mbox{\tiny$\begin{pmatrix}k\\l\end{pmatrix}$}{\phantom{\big|}}}$ &
$\sum_{p} \eta^{(\rho,\sig) M^p
  \mbox{\tiny$\begin{pmatrix} k \\ l \end{pmatrix}$}{\phantom{\big|}}}$ &
$0$ & $0$
\end{tabular}}
\end{center}

  \caption{\label{characterA} The character tables of $\Delta(3n^2)$
  for (a) $n \neq 3 \, \mathbb{Z}$ and (b) $n=3\,\mathbb{Z}$.
  Here $\om \equiv e^{\frac{2\pi
  \mathrm{i}}{3}}$ and $\eta \equiv e^{\frac{2\pi \mathrm{i}}{n}}$. The sum
  runs over $p=0,1,2$.}
\end{table}
\label{table}

\cleqn
\section{Kronecker Products}\label{product}

The construction of a theory that is invariant under a group of transformation requires the knowledge of its polynomial invariants. Simplest  are the quadratic invariants, which occur whenever the Kronecker product of two representations contains the singlet representation.  Given two irreps  $\bf{r}$ and $\bf s$,  their Kronecker product

\beqn \bf{r} \otimes \bf{s} & = & \sum_{\bf t}d(\bf
{r},\bf{s},\bf{t}) ~ \bf{t}\ , \eeqn can be expressed as a sum of
irreducible representations. The integer numerical factors $d(\bf
{r},\bf{s},\bf{t})$ can be calculated from the character table by
means of the formula

\beqn
d(\bf {r},\bf{s},\bf{t}) & = & \frac{1}{N} \sum_{i} n_i \cdot \chi^{[\bf r]}_i \, \chi^{[\bf s]}_i
\, \ol{\chi}^{[\bf t]}_i\ , \label{decomp} \eeqn
where $N$ is the order of the group; $i$ labels a class of $n_i$ elements and
character $\chi^{}_{i}$.   $\ol{\chi}_i$~denotes the complex conjugate
character. The sum over classes is rather intricate and, although the cases
$n \neq 3\, \mathbb{Z}$ and $n = 3\, \mathbb{Z}$ are to some extent similar, we discuss them separately.

\begin{itemize}

\item[$(i)$] {$n \neq  3 \, \mathbb{Z}$.}

The character table for this case is given in Table~\ref{characterA}a. There are three ($r=0,1,2$) one-dimensional
and $\frac{n^2-1}{3}$ three-dimensional irreducible representations. The sum over each class is complicated by the redundant $(k,l)$ labeling, and its  detailed calculation is left to Appendix~\ref{calcdetail}. Here we simply write the results

\beqn
 {\bf  1}_r \otimes {\bf 1}_{r^\prime} &=& {\bf 1}_{r + r^{\prime}} \notag \\
 {\bf 1}_r \otimes {\bf 3}_{(k,l)} &=& {\bf 3}_{(k,l)} \notag \\
{\bf  3}_{(k,l)} \otimes {\bf 3}_{(k^\prime,l^\prime) } &=&
\delta_{\mbox{\tiny$\begin{pmatrix}
k^\prime\\l^\prime\end{pmatrix}$},
            \widetilde{\mbox{\tiny$\begin{pmatrix} -k \\ -l \end{pmatrix}$}}}
~ ({\bf 1}_0  + {\bf 1}_1 +{\bf 1}_2) ~ + \notag \\&&\notag\\
&& \hspace{.1cm} +~{\bf 3}_{\widetilde{\mbox{\tiny$\begin{pmatrix} k^\prime+k \\
l^\prime+l \end{pmatrix}$}}}
+~{\bf 3}_{\widetilde{\mbox{\tiny$\begin{pmatrix} k^\prime-k-l \\
l^\prime+k  \end{pmatrix}$}}} +~
{\bf 3}_{\widetilde{\mbox{\tiny$\begin{pmatrix} k^\prime+l \\ l^\prime-k-l
\end{pmatrix}$}}}. \notag
\eeqn

\item[$(ii)$] {$n =  3 \, \mathbb{Z}$.}

The character table in this case is given in Table~\ref{characterA}b.
We have nine ($r,s=0,1,2$) one-dimensional and $\frac{n^2-3}{3}$
three-dimensional representations. Again the subtleties of evaluating the sum
are relegated to Appendix~\ref{calcdetail}. The result is

\beqn
{\bf 1}_{r,s} \otimes {\bf 1}_{r^\prime,s^\prime} &=&
 {\bf 1}_{r+r^\prime,s+s^\prime} \notag \\
{\bf 1}_{r,s} \otimes {\bf 3}_{(k,l) } &=&
  {\bf 3}_{\widetilde{\mbox{\tiny$\begin{pmatrix}k+{sn}/{3} \\
l+{sn}/{3}\end{pmatrix}$}}} \notag \\
{\bf 3}_{(k,l)} \otimes {\bf 3}_{(k^\prime,l^\prime) } &=& \sum_{s=0}^2
\delta^{}_{ \mbox{\tiny$\begin{pmatrix} k^\prime \\l^\prime
\end{pmatrix}$} , \widetilde{\mbox{\tiny$\begin{pmatrix} - k +
{sn}/{3}   \\  - l + {sn}/{3}
\end{pmatrix}$}}} \, ({\bf 1}_{0,s} + {\bf 1}_{1,s} + {\bf 1}_{2,s}) ~ + \notag \\
&&\notag\\
&&\hspace{.3cm}  + ~~ {\bf 3}_{\widetilde{\mbox{\tiny$\begin{pmatrix} k^\prime+k \\
l^\prime+l \end{pmatrix}$}}}
+~{\bf 3}_{\widetilde{\mbox{\tiny$\begin{pmatrix} k^\prime-k-l \\
l^\prime+k  \end{pmatrix}$}}} +~
{\bf 3}_{\widetilde{\mbox{\tiny$\begin{pmatrix} k^\prime+l \\ l^\prime-k-l
\end{pmatrix}$}}}. \notag
\eeqn

\end{itemize}

\cleqn
\section{Building Invariants}\label{invariants}
In the previous section we have decomposed the product of two
three-dimensional representations  in terms of irreducible
representations. In this section we use these results to explicitly
build invariants out of fields that transform as triplets. The
corresponding Clebsch-Gordan coefficients are derived in
Appendix~\ref{CGapp}. Throughout this section we use the following
notation: $\vec\varphi$ denotes a ${\bf 3}_{(k,l)}$ and
$\vec\varphi^\prime$ is a ${\bf 3}_{(k^\prime,l^\prime)}$. In terms
of the component fields we have

\beq {\bf 3}_{(k,l)}~=~ \vec\varphi ~=~ \begin{pmatrix} \varphi_1 \\ \varphi_2 \\
\varphi_3 \end{pmatrix}\ , ~~~~~~~~~~~~~~~{\bf  3}_{(k^\prime,l^\prime)}~=~
\vec\varphi^\prime ~=~
\begin{pmatrix} \varphi_1^\prime \\ \varphi_2^\prime \\ \varphi_3^\prime \end{pmatrix}\ ,
\eeq
and  the action of the generators in the triplet representation on these
fields is given by the matrices in Eq.~(\ref{3drep}).  We discuss separately the construction of  invariants, depending on whether $n$ is a multiple of three or not.


\begin{itemize}
\item[($i$)] {$n \neq  3 \, \mathbb{Z}$.}

In order to obtain a product of $\vec\varphi$ and
$\vec\varphi^\prime$ which transforms as one of the three
one-dimensional irreducible representations in Eq.~(\ref{33}) we
need to have

\beqn
\begin{pmatrix} k^\prime \\ l^\prime \end{pmatrix} &=&
\widetilde{\begin{pmatrix} -k \\ -l \end{pmatrix}}~=~ M^p
\begin{pmatrix} -k \\ -l \end{pmatrix},\label{arbitrary} \eeqn
for some value of $p=0,1,2$. From the character Table~\ref{characterA}a, we see that $\vec\varphi^\prime$ transforms as the complex conjugate representation of $\vec\varphi$.  The ambiguity of the $\widetilde{\phantom{va}}$ mapping yields three combinations that transform as the singlet ${\bf 1}_0$,

\beqn
p=0:&~~&\varphi_1^\prime \varphi_1 + \varphi_2^\prime \varphi_2 + \varphi_3^\prime \varphi_3,\notag\\
p=1:&~~&\varphi_1^\prime \varphi_2 + \varphi_2^\prime \varphi_3 +
\varphi_3^\prime\varphi_1,
\label{whatisp}  \\
p=2:&~~&\varphi_1^\prime \varphi_3 + \varphi_2^\prime \varphi_1 + \varphi_3^\prime
\varphi_2.\notag \eeqn
It is easy to verify the invariance of these expressions by acting the generating matrices of Eq.~(\ref{3drep}) on the fields.



For completeness, and mindful of the ambiguity Eq.~(\ref{arbitrary}), we choose the case $p=0$, and represent the three different one-dimensional representations
built out of  $\vec\varphi^\prime$ and $\vec\varphi$ as ($r=0,1,2$)

\beq {\bf 1}_r:~~~\vec\varphi^\prime \cdot
{\begin{pmatrix} 1 & 0 & 0 \\ 0 & \om^{-1}&0\\ 0&0&\om
\end{pmatrix}}^{\!\!r} \vec\varphi\,, ~~~~~~~~\mathrm{with}~~~~
\begin{pmatrix} k^\prime \\ l^\prime \end{pmatrix} =
{\begin{pmatrix} -k \\ -l \end{pmatrix}}.\label{2singlets}
\eeq
These are helpful in the construction of cubic invariants out of two triplets and one field that transforms as a one-dimensional representation.

Now we turn to the construction  of cubic invariants out of three triplets. For this we need  to build  triplets out of two triplets.  To be specific, let us focus on the second ${\bf 3}_{(k^{\prime\prime},l^{\prime\prime})}$ on the right-hand side of
Eq.~(\ref{33}).\footnote{The first ${\bf 3}_{(k^{\prime\prime},l^{\prime\prime})}$ is absent in the case
where  $(k^\prime,l^\prime)=(-k,-l)$.} Due to the mapping $\widetilde{\phantom{va}}$, we have

\beqn
\begin{pmatrix} k^{\prime\prime} \\ l^{\prime\prime} \end{pmatrix} &=& M^p
\begin{pmatrix} k^\prime -k-l\\ l^\prime +k \end{pmatrix},\label{kpplpp}
\eeqn
where $p=0,1,\,\mathrm{or}~2$.
We find three ways to build the ${\bf 3}_{(k^{\prime\prime},l^{\prime\prime})}$ out of the two triplets $\vec\varphi^\prime$ and $\vec\varphi$

\beq
{\bf 3}^{}_{(k^{\prime\prime},l^{\prime\prime})}:~~
\begin{pmatrix} \varphi_1^\prime \, \varphi_2 \\
                \varphi_2^\prime \, \varphi_3  \\
                 \varphi_3^\prime \, \varphi_1    \end{pmatrix}_{p=0},~~~~
\begin{pmatrix}  \varphi_2^\prime \, \varphi_3  \\
                 \varphi_3^\prime \, \varphi_1  \\
                 \varphi_1^\prime \, \varphi_2   \end{pmatrix}_{p=1}\ ,~~~~
\begin{pmatrix}  \varphi_3^\prime \, \varphi_1  \\
                 \varphi_1^\prime \, \varphi_2 \\
                  \varphi_2^\prime \, \varphi_3  \end{pmatrix}_{p=2}\ .\label{trip}
\eeq
%
These three triplets are related by a cyclic permutation of the components
expressible by the action of $a$.

The cubic invariant is obtained by multiplying the appropriate triplet in Eq.~(\ref{trip}) with the complex conjugate of a third  triplet
$\vec\varphi^{\prime\prime}$ which belongs to ${\bf 3}_{(k^{\prime\prime},l^{\prime\prime})}$, as in Eq.~(\ref{kpplpp}).

Denoting the complex conjugate field by $\ol{\varphi}_i$, the cubic invariants are given by

\beq
{\bf 1}_0:~~~\begin{pmatrix}  \bar{\varphi}^{\prime\prime}_1 &   \bar{\varphi}^{\prime\prime}_2 &
  \bar{\varphi}^{\prime\prime}_3 \end{pmatrix} \, a^p \,
\begin{pmatrix}
 \varphi_1^\prime  \,  \varphi_2 \\
\varphi_2^\prime  \,  \varphi_3 \\
 \varphi_3^\prime  \,  \varphi_1
\end{pmatrix}.
\eeq
This result depends on the value of $p$ in Eq.~(\ref{kpplpp})
and thus on the choice of the mapping~$\widetilde{\phantom{va}}$.
Analogous to Eq.~(\ref{2singlets}) we obtain all three singlets

\beqn
{\bf 1}_r:~~~\begin{pmatrix}  \bar{\varphi}^{\prime\prime}_1 &   \bar{\varphi}^{\prime\prime}_2 &
  \bar{\varphi}^{\prime\prime}_3 \end{pmatrix}
\, a^p \, {\begin{pmatrix} 1 & 0 & 0 \\ 0 & \om^{-1}&0\\
0&0&\om \end{pmatrix}}^{\!\!r} 
\begin{pmatrix}
 \varphi_1^\prime  \,  \varphi_2 \\
\varphi_2^\prime  \,  \varphi_3 \\
 \varphi_3^\prime  \,  \varphi_1
\end{pmatrix}\!\!,\label{3singlet2}
\eeqn
with
\beqn
\begin{pmatrix} k^{\prime\prime} \\ l^{\prime\prime} \end{pmatrix} \! =
M^p \begin{pmatrix} k^\prime -k-l\\ l^\prime +k \end{pmatrix}\!.~
\eeqn

So far we have only considered the second triplet in Eq.~(\ref{33}).
Alternatively, we can obtain one-dimensional representations from
the first and the third ${\bf
3}_{(k^{\prime\prime},l^{\prime\prime})}$ in Eq.~(\ref{33}) as well.
Generalizing Eq.~(\ref{kpplpp}) to include these possibilities leads
to

\beqn
\begin{pmatrix} k^{\prime\prime} \\ l^{\prime\prime} \end{pmatrix} &=& M^p
\left[ \begin{pmatrix} k^\prime \\ l^\prime \end{pmatrix} + M^q
\begin{pmatrix} k \\ l \end{pmatrix} \right]\ . \label{kpplppGEN}
\eeqn
$q=0$ corresponds to the first triplet of Eq.~(\ref{33}), $q=1$ to the second [hence it gives back Eq.~(\ref{kpplpp})], and
$q=2$ to the third. As above the value of $p$ depends on the choice of the standard representatives. With this notation, the cubic combinations that transform as the one-dimensional representations are given, in a way similar  to Eq.~(\ref{3singlet2}), by

\beq {\bf 1}_r:~~~~
\bar{\vec{\varphi}}^{\,\prime\prime} \cdot \, a^p \,
{\begin{pmatrix} \varphi_1^\prime & 0 & 0 \\ 0 & \om^{-r}\varphi_2^\prime &0 \\
    0&0&\om^r \varphi_3^\prime  \end{pmatrix}} \,  a^q \, \vec\varphi\ .
\label{3singlets} \eeq
The pairs $(k,l)$ and $(k^\prime,l^\prime)$ can take any (standard) value. $(k^{\prime\prime},l^{\prime\prime})$, which denotes the
representation of $\vec\varphi^{\,\prime\prime}$, is defined in Eq.~(\ref{kpplppGEN}). $q$
is 0, 1, {\it and} 2, while the value of $p$ is {\it either} 0 {\it or} 1 {\it or} 2, depending on the choice of the standard
representatives. As mentioned before, the first triplet in Eq.~(\ref{33}) is absent in the case where
$(k^\prime,l^\prime)=(-k,-l)$. Then the choice $q=0$ is forbidden in Eq.~(\ref{3singlets}).

\item[($ii$)] $n =  3 \cdot \mathbb{Z}$.

The decomposition of the product of two triplets in terms of irreducible
representations depends on whether or not the parameters
$k,l,k^\prime$, and $l^\prime$ are
multiples of $\frac{n}{3}$. If they are, one obtains either nine or zero
one-dimensional represetations; if at least one of the four parameters
is not a multiple of $\frac{n}{3}$, we instead
obtain three or zero
one-dimesional representations (see Appendix~\ref{calcdetail}). We discuss
each case separately.

\begin{itemize}

\item[$(a)$]$k,l,k',l'$ are  all multiples of $n/3$.

As discussed below Eq.~(\ref{33b}) the
pairs $(k,l)$ and $(k^\prime,l^\prime)$ can take only two different
values. Choosing $(0,\frac{n}{3})$ and
$(0,\frac{2n}{3})=(0,-\frac{n}{3})$ as the standard representatives,
we represent the nine different one-dimensional represenations built out of
$\vec\varphi^\prime$ and $\vec\varphi$ as $(r,s=0,1,2)$

\beq {\bf 1}_{r,s}:~~~
 \vec\varphi^\prime \cdot a^{\mp s} \,
{\begin{pmatrix} 1 & 0 & 0 \\ 0 & \om^{-1}&0\\ 0&0&\om
\end{pmatrix}}^{\!\!r} \, \vec\varphi\,, ~~~\mathrm{with}~
\begin{pmatrix} -k^\prime \\ -l^\prime \end{pmatrix} =
\begin{pmatrix} k \\ l \end{pmatrix} =
\begin{pmatrix} 0 \\ \pm n/3 \end{pmatrix}. \label{2singletsa}
\eeq

Except for the additional factor of $a^{\mp s}$, this expression is identical
to Eq.~(\ref{2singlets}). Under the action of $a \,/\, c,d$ it gets
multiplied by  $\om^r /\, \om^s$, respectively; that is, it
transforms as a ${\bf 1}_{r,s}$.


Let us now turn to the construction of the three-dimensional representations
out of two triplets $\vec\varphi^\prime$ and $\vec\varphi$. If
$(k^\prime,l^\prime)=(k,l)=(0,\pm\frac{n}{3})$,
we obtain three ${\bf 3}_{(k^{\prime\prime},l^{\prime\prime})}$
representations

\beq
{\bf 3}_{(k^{\prime\prime},l^{\prime\prime})}:~~~~
\begin{pmatrix}
\varphi^\prime_1 \, \varphi_1 \\
\varphi^\prime_2 \, \varphi_2 \\
\varphi^\prime_3 \, \varphi_3  \end{pmatrix},~~~~~~~~
\begin{pmatrix}
\varphi^\prime_2 \, \varphi_3 \\
\varphi^\prime_3 \, \varphi_1 \\
\varphi^\prime_1 \, \varphi_2  \end{pmatrix},~~~~~~~~
\begin{pmatrix}
\varphi^\prime_3 \, \varphi_2 \\
\varphi^\prime_1 \, \varphi_3 \\
\varphi^\prime_2 \, \varphi_1  \end{pmatrix},
\eeq

which can be written in
a more compact form, with $q=0,1,2$ labeling the three possibilities,

\beq
{\bf 3}_{(k^{\prime\prime},l^{\prime\prime})}:~~~~
a^q \begin{pmatrix}
\varphi^\prime_1 & 0 & 0 \\
0 & \varphi^\prime_2 & 0 \\
0 & 0 & \varphi^\prime_3 \end{pmatrix}  a^{q}
\begin{pmatrix}  \varphi_1 \\ \varphi_2 \\ \varphi_3  \end{pmatrix} .
\label{2tripletsb}
\eeq

Since these new constructed triplets transform as
$(k^{\prime\prime},l^{\prime\prime})=(0,\mp \frac{n}{3})$
representations, they constitute complex
conjugate representations of the initial triplets $\vec\varphi^\prime$ and
$\vec\varphi$.

The cubic invariants can be built by multiplying the three
($q=0,1,2$) triplets in Eq.~(\ref{2tripletsb}) with a third triplet
$\vec\varphi^{\prime\prime\prime}$ transforming as
$(k^{\prime\prime\prime},l^{\prime\prime\prime})=(0,\pm
\frac{n}{3})$. Using Eq.~(\ref{2singletsa}), we obtain the cubic
combinations that transform as the one-dimensional
representations\footnote{Note that we have specified the standard
representatives explicitly. Therefore a parameter like $p$ in
Eq.~(\ref{3singlets}) is not needed.}

\beqn {\bf 1}_{r,s}:&&
 \vec\varphi^{\prime\prime\prime} \cdot
 a^{\pm s} \,
a^q \begin{pmatrix}
\varphi^\prime_1 & 0 & 0 \\
0 & \om^{-r} \varphi^\prime_2 & 0 \\
0 & 0 & \om^{r} \varphi^\prime_3 \end{pmatrix}  a^{q} \, \vec\varphi,~~~
 \label{3singletsa}
\eeqn
with
\beqn
\begin{pmatrix} k^{\prime\prime\prime} \\ l^{\prime\prime\prime} \end{pmatrix} =
\begin{pmatrix} k^\prime \\ l^\prime \end{pmatrix} =
\begin{pmatrix} k \\ l \end{pmatrix} =
\begin{pmatrix} 0 \\ \pm n/3 \end{pmatrix}.
\eeqn

The parameter $q$ indicates that there are three different possibilities
to build the one-dimensional representation ${\bf 1}_{r,s}$ from the product
of $\vec\varphi^{\prime\prime\prime}$, $\vec\varphi^{\prime}$, and
$\vec\varphi$.

\item[$(b)$] $k,l,k',l'$ are not all multiples of $n/3$.

In order to obtain the three
possible one-dimensional representations by multiplying two
triplets $\vec\varphi^\prime$ and
$\vec\varphi$, we need to have

\beqn
\begin{pmatrix} k^\prime \\ l^\prime \end{pmatrix} & = &
\widetilde{\begin{pmatrix} -k + sn/3 \\ -l + sn/3 \end{pmatrix}} =
M^p \begin{pmatrix} -k  \\ -l  \end{pmatrix} + \frac{sn}{3}
\begin{pmatrix} 1  \\ 1  \end{pmatrix}.\label{kplps}
\eeqn

The value
of $s$ determines which set of singlets we are constructing; for example,
with $s=0$ we get the three singlets ${\bf 1}_{0,0}$, ${\bf 1}_{1,0}$, and
${\bf 1}_{2,0}$.
With Eq.~(\ref{kplps}), we can calculate the one-dimensional
representations similar to Eq.~(\ref{2singlets})

\beq {\bf 1}_{r,s}:~~~~~ \vec\varphi^\prime \cdot a^p \,
{\begin{pmatrix} 1 & 0 & 0 \\ 0 & \om^{-1}&0\\ 0&0&\om
\end{pmatrix}}^{\!\!r} \, \vec\varphi\,.\label{2singletsb}
\eeq

In this
expression, $r$ can take any of its three possible values, while $p$
is determined in Eq.~(\ref{kplps}) by
the values of $(k,l)$ and $s$ as well as the
mapping~$\widetilde{\phantom{va}}$.

The triplets obtained from the product of  $\vec\varphi^\prime$ and
$\vec\varphi$ are constructed
analogous to Eq.~(\ref{trip}). Multiplication with the appropriate
triplet $\vec\varphi^{\prime\prime}$ yields the cubic combinations that
transform as the one-dimensional representations

 \beqn {\bf 1}_{r,s}~&&
\vec\varphi^{\prime\prime} \cdot \, a^p \,
\begin{pmatrix}
\varphi^\prime_1 & 0 & 0 \\
0 & \om^{-r} \varphi^\prime_2 & 0 \\
0 & 0 & \om^{r} \varphi^\prime_3 \end{pmatrix}
\, a^q \, \vec\varphi, \label{3singletsb}
\eeqn
with
\beqn
\begin{pmatrix} k^{\prime\prime} \\ l^{\prime\prime} \end{pmatrix} =
M^p  \left[   -  \begin{pmatrix} k^{\prime} \\ l^{\prime}
\end{pmatrix}
 - M^q    \begin{pmatrix} k \\ l \end{pmatrix}   \right] +
\frac{sn}{3}  \begin{pmatrix} 1 \\ 1 \end{pmatrix}.
\eeqn

The pairs $(k,l)$ and $(k^\prime,l^\prime)$ can take any (standard)
value. $s$ and $r$ determine what type of singlet we want to build.
$q$ can be 0, 1, and 2, while $p$ is restricted to only one value
[depending on the choice of the standard representative for
$(k^{\prime\prime},l^{\prime\prime})$]. In the case where the
product of $\vec\varphi^\prime$ and $\vec\varphi$ already includes
singlets, i.e. where $\begin{pmatrix} k^\prime \\ l^\prime
\end{pmatrix} = M^{p^\prime} \begin{pmatrix} -k + s^\prime n/3 \\ -l
+ s^\prime n/3
\end{pmatrix}$ for some pair $p^\prime,s^\prime=0,1,2$, we must omit
the choice $q=p^\prime$ in Eq.~(\ref{3singletsb}).
\end{itemize}
\end{itemize}
In summary, the products of two $\Delta(3n^2)$ triplet fields
yielding one-dimensional representations are given in
Eqs.~(\ref{2singlets}), (\ref{2singletsa}), and (\ref{2singletsb}).
The products of three triplets which - as a product - transform as
one-dimensional representations are stated in
Eqs.~(\ref{3singlets}), (\ref{3singletsa}), and (\ref{3singletsb}).

\cleqn
\section{Outlook}

In this article we have studied the class structure of the group
$\Delta(3n^2)$ and its irreducible representations. We have derived the
Kronecker products and constructed the Clebsch-Gordan coefficients. In order
to formulate a $\Delta(3n^2)$-invariant theory it is necessary to know how to
build the (quadratic and cubic) invariants. Previous studies usually restrict
themselves to building invariants in a specific way. For example,
the authors of Refs.~\cite{Ross} only consider products of a triplet ${\bf
  3}_{(k,l)}$ of $\Delta(27)$ and its complex conjugate ${\bf
  \bar{3}}_{k,l}={\bf 3}_{(-k,-l)}$; also they do not include particles
transforming as ${\bf 1}_{r,s}$ representations with $r,s \neq 0$.
Other models of flavor (e.g. \cite{Ma}) adopt less trivial ways to
build invariants. However, a systematic presentation of ``what is
possible'' is still missing. Here we have tried to fill this gap. We
hope that our study will be useful for model builders aiming at a
thorough investigation of flavor models based on the group
$\Delta(3n^2)$. In future publication, we plan to apply the same
techniques to $\Delta(6n^2)$ and other finite groups such as
$PSL_2(7)$.

\section*{Acknowledgments}

The work of CL is supported by the University of Florida through the
Institute for Fundamental Theory and that of SN and PR is supported
by the Department Of Energy Grant No. DE-FG02-97ER41029.

\section*{Appendix}

\appendix


\section{Labeling the Three-Dimensional Representations}
\label{standard}

When labeling the three-dimensional representations of $\Delta(3n^2)$
by the subscripts $(k,l)$
there is a certain ambiguity involved. Taking, say, $(0,1)$ is
equivalent to the pairs $(-1,0)$ and $(1,-1)$ [see Eq.~(\ref{3pairs})]. It is
crucial for the discussion of the products of two representations that we list
only one of the three equivalent representations. For this purpose the mapping
$\widetilde{\phantom{va}}$  was introduced in Eq.~(\ref{mapping}). It remained
unspecified in Section~\ref{sirreps}. In this appendix we wish to give one
possible definition of $\widetilde{\phantom{va}}$ by picturing the full set of
{\it standard} representatives for $(k,l)$ on the $n \times n$ grid. We have to
distinguish between three cases: $(i_1)$ $n=3z + 1$ with $z\in \mathbb{Z}$,
$(i_2)$ $n=3z + 2$, and $(ii)$ $n=3z$.

\begin{itemize}
\item[($i_1$)] $n = 3 z +1$.\vspace{-4mm}
\begin{center}
\begin{tabular}{c||c|cccccccccccc}
$(k,l)$ & 0 &1& $\cdot$ &$\cdot$ & $z$ &$\cdot$ &$\cdot$ &$\cdot$ &$2z$
&$\cdot$ & $\cdot$ &$\cdot$  & $3z$ \\ \hline\hline
0& & $\cdot$ & $\cdot$ &$\cdot$ &  $\cdot$ &  $\cdot$ &$\cdot$ &$\cdot$
&$\cdot$ &  $\cdot$ &$\cdot$ &  $\cdot$ & $\cdot$\\ \hline
1& & $\cdot$ & $\cdot$ &$\cdot$ &  $\cdot$ &  $\cdot$ &$\cdot$ & $\cdot$ & &   & &  &\\
$\cdot$ & & $\cdot$ & $\cdot$ &$\cdot$ &  $\cdot$ &  $\cdot$ & $\cdot$ &
& &   & &  &\\
$\cdot$ & & $\cdot$ & $\cdot$ &$\cdot$ &  $\cdot$ & $\cdot$ & &
& &   & &  &\\
$z$& &$\cdot$ &  $\cdot$ & $\cdot$ & $\cdot$ & & & &   & &  &&\\
$\cdot$ &&&&&&&&&&&&& \\
$\cdot$ &&&&&&&&&&&&& \\
$\cdot$ &&&&&&&&&&&&& \\
$2z$ & &&&&&&&&& & & &  \\
$\cdot$&&&&&&&&&& $\cdot$ & $\cdot$ & $\cdot$ & $\cdot$ \\
$\cdot$&&&&&&&&&$\cdot$& $\cdot$ & $\cdot$ & $\cdot$ & $\cdot$ \\
$\cdot$&&&&&&&&$\cdot$&$\cdot$& $\cdot$ & $\cdot$ & $\cdot$ & $\cdot$ \\
$3z$&&&&&&&$\cdot$&$\cdot$&$\cdot$&$\cdot$&$\cdot$&$\cdot$&$\cdot$\\
\end{tabular}
\end{center}
The dots indicate the chosen standard representatives for $(k,l)$.
The diagram consists of three regions. First there are the representations
with $k=0$ and $l=1,2,...,3z$. This results in $3z$ representations.
Note that $(k,l)=(0,0)$ does not label an
irreducible representation. The second region is in the upper left
corner. Commencing with the $z \times z$ square, we extend the area of
representations to the right, reducing the value of the largest
$k$ by one in each
step: For $l=z+1$ we have $k\leq z-1$, then $l=z+2$ requires $k\leq z-2$, etc.
Thus we obtain $z^2 + \frac{z(z-1)}{2}$ representations in the second
region. The third region is obtained from the second by taking the negative
values of the pairs $(k,l)$ in the second region and shifting these by
$n$. For instance,
$(1,1)$ of the second region is transferred to $(-1,-1) = (3z,3z)$.
Therefore the representations in the third region are the complex conjugates
of those in the second region.\\
Altogether we obtain $3z+2z^2+z(z-1) = \frac{1}{3}[(9z^2 + 6z +1)-1]
= \frac{n^2-1}{3}$ representations, a value which is required for $n\neq
3 \, \mathbb{Z}$.  It can be shown that all of these representations
are inequivalent.

\item[($i_2$)] $n = 3 z +2$.\vspace{-4mm}
\begin{center}
\begin{tabular}{c||c|ccccccccccccl}
$(k,l)$ & 0 &1& $\cdot$ &$\cdot$ & $z$ &$\cdot$ &$\cdot$ &$\cdot$ &$2z$
&$\cdot$ & $\cdot$ &$\cdot$  & $3z$ & $\!\!\!\!3z\!\!+\!\!1$ \\ \hline\hline
0& & $\cdot$ & $\cdot$ &$\cdot$ &  $\cdot$ &  $\cdot$ &$\cdot$ &$\cdot$
&$\cdot$ &  $\cdot$ &$\cdot$ &  $\cdot$ & $\cdot$& $\cdot$\\ \hline
1& & $\cdot$ & $\cdot$ &$\cdot$ &  $\cdot$ &  $\cdot$ &$\cdot$ & $\cdot$&
$\cdot$  & &   & &  &\\
$\cdot$ & & $\cdot$ & $\cdot$ &$\cdot$ &  $\cdot$ &  $\cdot$ & $\cdot$
&$\cdot$ & & &   & &  &\\
$\cdot$ & & $\cdot$ & $\cdot$ &$\cdot$ &  $\cdot$ & $\cdot$ &$\cdot$ & &
& &   & &  &\\
$z$& &$\cdot$ &  $\cdot$ & $\cdot$ & $\cdot$ &$\cdot$ & & & &   & &  &&\\
$\cdot$ &&&&&&&&&&&&& \\
$\cdot$ &&&&&&&&&&&&& \\
$\cdot$ &&&&&&&&&&&&& \\
$2z$ & &&&&&&&&& & & &  \\
$\cdot$ & &&&&&&&&& & & &  \\
$\cdot$&&&&&&&&&& $\cdot$& $\cdot$ & $\cdot$ & $\cdot$ & $\cdot$ \\
$\cdot$&&&&&&&&& $\cdot$&$\cdot$& $\cdot$ & $\cdot$ & $\cdot$ & $\cdot$ \\
$3z$&&&&&&&& $\cdot$&$\cdot$&$\cdot$& $\cdot$ & $\cdot$ & $\cdot$ & $\cdot$ \\
$3z+1$&&&&&&&$\cdot$&$\cdot$&$\cdot$&$\cdot$&$\cdot$&$\cdot$&$\cdot$&$\cdot$\\
\end{tabular}
\end{center}
In this case the table of standard representatives for $(k,l)$ is
obtained similarly to the above. We have $3z+1$ pairs in the first
region. Compared to $(i_1)$, the triangle on the right side of the
second region is shifted one step to the right and the free column
is filled up with $z$ representations. This gives
$z^2+\frac{(z+1)z}{2}$ pairs. The third region is again only the
complex conjugate of the second one. Summing the number of all
representations yields the required value of $\frac{n^2-1}{3}$.
%
%

\item[($ii$)] $n = 3 z $.\vspace{-4mm}
\begin{center}
\begin{tabular}{c||c|ccccccccccl}
$(k,l)$ & 0 &1& $\cdot$ &$\cdot$ & $z$ &$\cdot$ &$\cdot$ &$\cdot$ &$2z$
&$\cdot$ & $\;\,\cdot\;\,$ & $\!\!\!\!3z\!\!-\!\!1$ \\ \hline\hline
0& &  $\cdot$ &$\cdot$ &  $\cdot$ &  $\cdot$ &$\cdot$ &$\cdot$
&$\cdot$ &  $\cdot$ &$\cdot$ &  $\cdot$ & $\cdot$\\ \hline
1& &  $\cdot$ &$\cdot$ &  $\cdot$ &  $\cdot$ &$\cdot$ & $\cdot$ & &   & &  &\\
$\cdot$ & &  $\cdot$ &$\cdot$ &  $\cdot$ &  $\cdot$ & $\cdot$ &
& &   & &  &\\
$\cdot$ & & $\cdot$ & $\cdot$ &  $\cdot$ & $\cdot$ & &
& &   & &  &\\
$z$& &$\cdot$ &   $\cdot$ & $\cdot$ & & & &   & &  &&\\
$\cdot$ &&&&&&&&&&&& \\
$\cdot$ &&&&&&&&&&&& \\
$\cdot$ &&&&&&&&&&&& \\
$2z$ & &&&&&&&& & $\cdot$ & $\cdot$ & $\cdot$ \\
$\cdot$&&&&&&&&&  $\cdot$ & $\cdot$ & $\cdot$ & $\cdot$ \\
$\cdot$&&&&&&&&$\cdot$& $\cdot$ & $\cdot$ & $\cdot$ & $\cdot$ \\
$3z-1$&&&&&&&$\cdot$&$\cdot$&$\cdot$&$\cdot$&$\cdot$&$\cdot$\\
\end{tabular}
\end{center}
Compared to the case $(i_1)$,
the triangle of the second region is shifted by one step to the left.
Because $(k,l)=(z,z)$ [and also $(2z,2z)$] does not give an
irreducible representation,
this point of the $z \times z$ square has to be excluded.
Counting all pairs of $(k,l)$ gives $\frac{n^2-3}{3}$
different representations as required for the case where $n=3 \, \mathbb{Z}$.
\end{itemize}

\section{The Method of Induced Representations}\label{inducedREP}
The method of induced representations is sometimes useful for building irreducible representations, although in most cases it yields reducible representations. Luckily,  it allows us to build the $\Delta(3n^2)$ triplet representations in a particularly simple way. Here we sketch the main ideas and apply them to our group.

Let $\m H$ be a subgroup of $\m G$ of index $N$, (order of $\m G$ divided by
that of $\m H$)

\beqn\m G~\supset~ \m H .\eeqn
Suppose we know a $d$-dimensional representation $\bf{r}$ of $\m H$, that is for every
$h\in \m H$,

\beqn h:~\ket i~\mapsto~ \m{M}(h)^{\bf{[ r]}}_{ji}\ket j\ ,\eeqn
acting on the $d$-dimensional Hilbert space $\mathfrak H$, spanned
by $\ket i$, $i=1,2,\dots, d$. This enables us to construct the {\em
induced representation} of $\m G$, generated by  the representation
of $\m H$. We begin by considering the coset made up of $N$ points,

\beqn
(\,\m G/\m H\,):~~~~\m H~\oplus~ g^{}_1\m H~\oplus~ g^{}_2\m
H~\oplus~\cdots~\oplus ~g^{}_{N-1}\m H\ ,
\eeqn
where $g_k$ ($k=1,...,N-1$) are elements of $\m G$ not in $\m H$.  Consider the set of elements

\beqn
\m G\times\mathfrak H:~~~(\,g\,,\,\ket i\,)\ ;\qquad  g~\in~ \m G\
,~\ket i~\in~ \mathfrak H\ .
\eeqn
In order to establish a one-to-one correspondence with the coset, we assume that whenever the $\m G$ group element is of the form

\beqn
g~=~g^{}_kh^{}_a\ ,
\eeqn
we make the identification

\beqn
(\,g^{}_kh^{}_a\,,\,\ket i\,)~=~(\,g^{}_k\,,\,\m
M(h^{}_a)^{\bf[r]}_{ji}\ket j\,)\ ,
\eeqn
so that both $g$ and $gh$ are equivalent in the sense that they differ only by reshuffling $\mathfrak
H$. In this way, we obtain $N$ copies of $\mathfrak H$, the Hilbert space of the $\bf r$ representation of $\m H$, one at each point of
the coset, which we take to be

\beqn (\,g_0 \equiv e\,,\,\mathfrak H\,)\ ,~(\,g^{}_1\,,\,\mathfrak H\,)\
,~(\,g^{}_2\,,\,\mathfrak H\,)\ ,~\cdots\ ,~(\,g^{}_{N-1}\,,\,\mathfrak
H\,)\ .\eeqn
The action of $\m G$ on this set is simply group multiplication

\beqn
  g:~~~(\,g^{}_k\,,\,\ket i\,) ~\mapsto~ (\,gg^{}_k\,,\,\ket i\,)\
,\qquad g~\in \m G\ ,~~ k=0,1,\cdots, N-1\ .\eeqn
 Suppose that $g=h_a$, the coset decomposition tells us that

\beqn
h^{}_ag^{}_k=g{}_lh^{}_b\ ,
\eeqn
where $g_l$ and $h_b$ are uniquely determined. Hence,

\beqn
h^{}_a:~~~(\,g^{}_k\,,\,\ket
i\,)~\mapsto~(\,h^{}_ag^{}_k\,,\,\ket
i\,)~=~(\,g^{}_lh^{}_b\,,\,\ket i\,)~=~(\,g^{}_l\,,\,\m
M(h_b)^{\bf[r]}_{ji}\ket j\,)\ .
\eeqn
Similarly, when $g=g_l$,  the product $g_lg_k$ is itself a group element and can be rewritten uniquely as

\beqn
g^{}_lg^{}_k~=~g^{}_mh^{}_c\ ,
\eeqn
for some $g_m$ and $h_c$. It follows that

\beqn
g^{}_l:~~~(\,g^{}_k\,,\,\ket
i\,)~\mapsto~(\,g^{}_lg^{}_k\,,\,\ket
i\,)~=~(\,g^{}_mh^{}_c\,,\,\ket i\,)~=~(\,g^{}_m\,,\,\m
M(h_c)^{\bf[r]}_{ji}\ket j\,)\ .
\eeqn
We have shown that these $N$ copies of the vector space are linearly mapped into one another under the action of $\m G$. The action of $\m G$ is thus represented by a $(dN\times dN)$ matrix, which is not necessarily  irreducible.

We now apply this method to  $\m G=\Delta(3n^2)$ and take the
subgroup to be $\m H=(\m Z_n\times\m Z_n)$, of index $N=3$. Since
the subgroup is Abelian, its representations are one-dimensional,
with generators\footnote{The exponents of $c$ and $d$ are chosen so
that they give the $(1,1)$-entries of the $(3 \times 3)$ matrices
$c$ and $d$ in Eq.~(\ref{3drep}).
 $\eta$ is the $n$-th root of unity: $ \eta^n~=~1$.}

\beqn c~=~\eta^l_{}\ ;\qquad d~=~\eta^{-k-l}_{}\  , \eeqn acting on
complex numbers $z$. The coset contains just three points, and we
consider the action of $\Delta(3n^2)$ on

\beqn
(\,e\,,\, z\,)\ ,\qquad (\,a^2_{}\,,\, z\,)\ ,\qquad (\,a_{}\,,\, z\,)\ .
\eeqn
Then

\beqn
a(\,e\,,\, z\,)=(\,a\,,\, z\,)\ ,\qquad a(\,a^2_{}\,,\,
z\,)=(\,e\,,\, z\,)\ ,\qquad a(\,a\,,\, z\,)=(\,a^2_{}\,,\,
z\,)\ ,
\eeqn
so that $a$ is represented by the $(3\times 3)$ permutation matrix

\beqn
a~=~\begin{pmatrix} 0&1&0\cr  0&0&1\cr1&0&0 \end{pmatrix}\ .
\eeqn
Next, comes the action of $c$

\beqn
 c(\,e\,,\, z\,)~=~(\,c\,,\, z\,)~=~(\,e\,,\, cz\,)~=~(\,e\,,\,
\eta^l z\,)~=~\eta^l(\,e\,,\, z\,)\ .
\eeqn
Using

\beqn
ca^2_{}~=~a^2_{}c^{-1}_{}d^{-1}_{}   \ ;\qquad   ca~=~ad  \ ,
\eeqn
derived from the presentation, we find

$$ c(\,a^2_{}\,,\, z\,)\:=\:(\,ca^2_{}\,,\,
z\,)\:=\:(\,a^2_{}c^{-1}_{}d^{-1}_{}\,,\, z\,)\:=\:(\,a^2_{}\,,\,
c^{-1}_{}d^{-1}_{} z\,)\:=\:(\,a^2\,,\, \eta^{-l}
\eta^{k+l}z\,)\:=\: \eta^{k}(\,a^2_{}\,,\, z\,)\, ,
$$
$$
c(\,a\,,\, z\,)~=~(\,ca\,,\, z\,)~=~(\,ad\,,\, z\,)~=~(\,a\,,\, d
z\,)~=~(\,a\,,\, \eta^{-k-l} z\,)~=~\eta^{-k-l}(\,a\,,\, z\,)\ .
$$
The action of $d$ is derived in the same way,  using $da^2_{}\,=\,a^2_{}c\,$,
$da\,=\,ac^{-1}_{}d^{-1}_{}$.
As a result, both $c$ and $d$ are represented by the diagonal matrices

\beqn c~=~\left(
  \begin{array}{ccc}
    \eta^l & 0 & 0 \\
    0 & \eta^k & 0\\
    0 & 0 & \eta^{-k-l}\\
  \end{array}
\right) ,\qquad d~=~\left(
  \begin{array}{ccc}
    \eta^{-k-l} & 0 & 0 \\
    0 & \eta^{l} & 0\\
    0 & 0 & \eta^k \\
  \end{array}
\right)\ ,\eeqn
which completes the construction of the induced $(3\times 3)$ representation.





\cleqn
\section{Product of Irreducible Representations: Details}\label{calcdetail}

Using Eq.~(\ref{decomp}) and the character
Tables~\ref{characterA} we calculate the numerical factors
$d({\bf r},{\bf s},{\bf  t})$ in this appendix.

\begin{itemize}

\item[$(i)$]  {$n \neq  3 \cdot \mathbb{Z}$.}

There are $\frac{n^2-1}{3}$ three-dimensional irreducible representations. We
need to exclude $(k,l)=(0,0)$, and we implicitly assume that $(k,l)$ only take their standard
values (see e.g. Appendix~\ref{standard}). In addition, for the classes
$C_1^{(\rho,\sig)}$, we need
to exclude $(\rho,\sig) = (0,0)$. Recalling that summing over the
remaining pairs $(\rho,\sig)$ overcounts
each class three times, we can write the sum over the classes $C_1^{(\rho,\sig)}$ formally as

\beqn
\sum_{C_1^{(\rho,\sig)}} & = & \frac{1}{3} \left(
\sum_{\rho,\sig=0}^{n-1}
  ~~~-~~~  \sum_{\rho=\sig=0} \right).\label{c1sum}
\eeqn

Noting that the characters of the class $C_1$  are identical
to the characters of $C_1^{(\rho,\sig)}$ with $(\rho,\sig) = (0,0)$,
we can write

\beqn \chi^{[{\bf r}]}_{C_1} \,
\chi^{[{\bf s}]}_{C_1} \, \ol{\chi}^{[{\bf t}]}_{C_1}  ~+
\sum_{C_1^{(\rho,\sig)}} 3 \cdot \chi^{[{\bf r}]}_{C_1^{(\rho,\sig)}}
\, \chi^{[{\bf s}]}_{C_1^{(\rho,\sig)}} \,
\ol{\chi}^{[{\bf t}]}_{C_1^{(\rho,\sig)}} &=&
\sum_{\rho,\sig=0}^{n-1}   \chi^{[{\bf r}]}_{C_1^{(\rho,\sig)}} \,
\chi^{[{\bf s}]}_{C_1^{(\rho,\sig)}} \,
\ol{\chi}^{[{\bf t}]}_{C_1^{(\rho,\sig)}}.~~~~~~~
\eeqn

Therefore, in Eq.~(\ref{decomp}), the sum
over the classes $C_1$ and $C_1^{(\rho,\sig)}$
can be neatly written as a sum over all $n^2$ pairs of $(\rho,\sig)$. With this
observation, we now calculate the products ${\bf 1}_r \otimes
{\bf 1}_{r^\prime}$, ${\bf 1}_r \otimes {\bf 3}_{(k,l)}$, and ${\bf 3}_{(k,l)} \otimes
 {\bf 3}_{(k^\prime,l^\prime)}$ in turn.
\begin{itemize}
\item[$\bullet$] ${\bf 1}_r \otimes {\bf 1}_{r^\prime}$.
\beqn d({\bf 1}_r,{\bf 1}_{r^\prime},{\bf 1}_{r^{\prime\prime}}) & = & \frac{1}{3n^2}
\cdot \left[ \left(\sum_{\rho,\sig=0}^{n-1}  1 \right)  + n^2 \cdot
\om^{r+r^\prime - r^{\prime\prime}} +
n^2 \cdot \om^{-(r+r^\prime -   r^{\prime\prime})}  \right]\notag \\
&=& \frac{1}{3} \cdot \left[ 1+  \om^{r+r^\prime - r^{\prime\prime}}
+ \om^{-(r+r^\prime -
    r^{\prime\prime})}  \right] \notag \\
&=&  \delta^{(3)}_{r^{\prime\prime},r+r^\prime}.\label{111} \eeqn
\beqn d({\bf 1}_r,{\bf 1}_{r^\prime},{\bf 3}_{(k,l)}) & = & \frac{1}{3n^2} \cdot
\sum_{\rho,\sig=0}^{n-1} \sum_{p=0}^2 \eta^{-(\rho,\sig) M^p
  \mbox{\tiny$\begin{pmatrix} k \\ l \end{pmatrix}$}} ~=~ 0.\label{113}
\eeqn Due to the relation $1+\om + \om^{-1} = 0$, Eq.~(\ref{111}) is
non-zero only if the exponent of $\om$ is zero mod\,(3). The
superscript $^{(3)}$ on the Kronecker delta indicates that we
calculate modulo~3. Analogously, due to $\sum_{\rho=0}^{n-1}
\eta^{\rho} = 0$, Eq.~(\ref{113}) would be non-zero only if
$(k,l)=(0,0)$. This choice however is forbidden. We thus have \beqn
{\bf 1}_r \otimes {\bf 1}_{r^\prime} & = & {\bf 1}_{r + r^{\prime}}. \eeqn
\item[$\bullet$] ${\bf 1}_r \otimes {\bf 3}_{(k,l)}$.
\beqn d({\bf 1}_r,{\bf 3}_{(k,l)},{\bf 1}_{r^\prime}) & = &\frac{1}{3n^2} \cdot
\sum_{\rho,\sig=0}^{n-1} \sum_{p=0}^2 \eta^{(\rho,\sig) M^p
\mbox{\tiny$\begin{pmatrix} k \\ l \end{pmatrix}$}} ~=~
0.\label{131} \eeqn \beqn d({\bf 1}_r,{\bf 3}_{(k,l)},{\bf 3}_{(k^\prime,l^\prime)}) &
=& \frac{1}{3n^2} \cdot \sum_{\rho,\sig=0}^{n-1}
\sum_{~p,p^\prime=0~}^2 \eta^{(\rho,\sig)
\left[M^p\mbox{\tiny$\begin{pmatrix} k \\ l \end{pmatrix}$}
-M^{p^\prime}\mbox{\tiny$\begin{pmatrix} k^\prime \\ l^\prime
  \end{pmatrix}$}\right]} \notag \\
&=& \frac{1}{n^2}\cdot \sum_{\rho,\sig=0}^{n-1} \sum_{p=0}^2
\eta^{(\rho,\sig)  \left[M^p\mbox{\tiny$\begin{pmatrix} k \\ l
\end{pmatrix}$} -\mbox{\tiny$\begin{pmatrix} k^\prime \\ l^\prime
  \end{pmatrix}$}\right]} \notag \\
&=&  \delta_{(k^\prime , l^\prime),{(k,l)}}.\label{133} \eeqn In the
first step of Eq.~(\ref{133}) we have absorbed $M^{p^\prime}$ into a
redefinition of $(\rho,\sig)$ and $p$, leading to a factor of 3. The
resulting expression only contributes a non-zero value if
$\begin{pmatrix} k^\prime \\ l^\prime
\end{pmatrix} = M^p \begin{pmatrix} k \\ l  \end{pmatrix}$. As
$(k^\prime,l^\prime)$ and
 $(k,l)$ take their standard values, this condition is satisfied
only for $p=0$ and thus $(k^\prime,l^\prime)=(k,l)$. Hence we find
the $r$ independent equation \beqn {\bf 1}_r \otimes {\bf 3}_{(k,l)} &=&
{\bf 3}_{(k,l)}. \eeqn
\item[$\bullet$] ${\bf 3}_{(k,l)} \otimes {\bf 3}_{(k^\prime,l^\prime) }$.
\beqn d({\bf 3}_{(k,l)},{\bf 3}_{(k^\prime,l^\prime)},{\bf 1}_{r}) & = &\frac{1}{3n^2}
\cdot \sum_{\rho,\sig=0}^{n-1} \sum_{~p,p^\prime=~0}^2
\eta^{(\rho,\sig) \left[ M^p \mbox{\tiny$\begin{pmatrix} k \\ l
\end{pmatrix}$} +
 M^{p^\prime} \mbox{\tiny$\begin{pmatrix} k^\prime \\ l^\prime \end{pmatrix}$}
\right]}  \notag \\
& = & \frac{1}{n^2}  \cdot \sum_{\rho,\sig=0}^{n-1} \sum_{p=0}^2
\eta^{(\rho,\sig) \left[ M^p \mbox{\tiny$\begin{pmatrix} k \\ l
\end{pmatrix}$} +
  \mbox{\tiny$\begin{pmatrix} k^\prime \\ l^\prime \end{pmatrix}$}
\right]} \notag \\
&=& \delta_{\mbox{\tiny$\begin{pmatrix}
k^\prime\\l^\prime\end{pmatrix}$},
\widetilde{\mbox{\tiny$\begin{pmatrix} -k \\ -l
\end{pmatrix}$}}}.\label{331} \eeqn For a non-zero contribution we
need $\begin{pmatrix} k^\prime \\ l^\prime
\end{pmatrix} = M^p \begin{pmatrix} -k \\ -l  \end{pmatrix}$. As
$(k^\prime,l^\prime)$ takes its standard value, only one value of
$p$ can satisfy this equation, namely the one that maps $(-k,-l)$
onto $\widetilde{(-k,-l)}$. \beqn
d({\bf 3}_{(k,l)},{\bf 3}_{(k^\prime,l^\prime)},{\bf 3}_{(k^{\prime\prime},l^{\prime\prime})})
& = & \frac{1}{3n^2}\cdot \sum_{\rho,\sig=0}^{n-1}
\sum_{~p,p^\prime,p^{\prime\prime}=~0}^2 \eta^{(\rho,\sig) \left[
M^p \mbox{\tiny$\begin{pmatrix} k \\ l \end{pmatrix}$} +
 M^{p^\prime} \mbox{\tiny$\begin{pmatrix} k^\prime \\ l^\prime \end{pmatrix}$}  -
 M^{p^{\prime\prime}} \mbox{\tiny$\begin{pmatrix} k^{\prime\prime} \\
     l^{\prime\prime}  \end{pmatrix}$}\right]} \notag \\
& = &\frac{1}{n^2} \cdot \sum_{\rho,\sig=0}^{n-1}
\sum_{~p,p^\prime=~0}^2 \eta^{(\rho,\sig) \left[ M^p
\mbox{\tiny$\begin{pmatrix} k \\ l \end{pmatrix}$} +
 M^{p^\prime} \mbox{\tiny$\begin{pmatrix} k^\prime \\ l^\prime \end{pmatrix}$}  -
  \mbox{\tiny$\begin{pmatrix} k^{\prime\prime} \\
     l^{\prime\prime}  \end{pmatrix}$}\right]} \notag \\
& = &\frac{1}{n^2} \cdot \sum_{\rho,\sig=0}^{n-1}
\sum_{~p,p^\prime=~0}^2 \eta^{(\rho,\sig) \left[  M^{p^\prime}
\left\{ M^p \mbox{\tiny$\begin{pmatrix} k \\ l \end{pmatrix}$} +
\mbox{\tiny$\begin{pmatrix} k^\prime \\ l^\prime \end{pmatrix}$}
\right\} - \mbox{\tiny$\begin{pmatrix} k^{\prime\prime} \\
l^{\prime\prime}\end{pmatrix}$}\right]}
\notag \\
& = & \delta_{\mbox{\tiny$\begin{pmatrix}
k^{\prime\prime}\\l^{\prime\prime}\end{pmatrix}$},
\widetilde{\mbox{\tiny$\begin{pmatrix} k^\prime+k \\ l^\prime+l
\end{pmatrix}$}} } + \delta_{\mbox{\tiny$\begin{pmatrix}
k^{\prime\prime}\\l^{\prime\prime}\end{pmatrix}$},
\widetilde{\mbox{\tiny$\begin{pmatrix} k^\prime-k-l \\ l^\prime+k
\end{pmatrix}$}} } + \delta_{\mbox{\tiny$\begin{pmatrix}
k^{\prime\prime}\\l^{\prime\prime}\end{pmatrix}$},
\widetilde{\mbox{\tiny$\begin{pmatrix} k^\prime+l \\ l^\prime-k-l
\end{pmatrix}$}} } .\notag \\\label{333} \eeqn Again, only one value
of $p^\prime$ contributes to this sum. On the other hand, the three
possibilities for $p$ persist, so that we are left with three
Kronecker deltas. From Eqs.~(\ref{331}) and (\ref{333}) we get \beqn
{\bf 3}_{(k,l)} \otimes {\bf 3}_{(k^\prime,l^\prime)} &=&
\delta_{\mbox{\tiny$\begin{pmatrix}
k^\prime\\l^\prime\end{pmatrix}$},
            \widetilde{\mbox{\tiny$\begin{pmatrix} -k \\ -l \end{pmatrix}$}}}
 ~ \cdot ~ ({\bf 1}_0  + {\bf 1}_1 +{\bf 1}_2) ~ + \notag \\
&&+~{\bf 3}_{\widetilde{\mbox{\tiny$\begin{pmatrix} k^\prime+k \\
l^\prime+l \end{pmatrix}$}}}
+~{\bf 3}_{\widetilde{\mbox{\tiny$\begin{pmatrix} k^\prime-k-l \\
l^\prime+k  \end{pmatrix}$}}} +~
{\bf 3}_{\widetilde{\mbox{\tiny$\begin{pmatrix} k^\prime+l \\ l^\prime-k-l
\end{pmatrix}$}}}.\label{33} \eeqn Notice that if
$(k^\prime,l^\prime) = \widetilde{(-k,-l)}$, then one of the three
{\bf 3}'s is absent as there is no representation ${\bf 3}_{(0,0)}$.

\end{itemize}

\item[$(ii)$]  {$n =  3 \, \mathbb{Z}$.}

In this case,
$(k,l)=(0,0),(\frac{n}{3},\frac{n}{3}),(\frac{2n}{3},\frac{2n}{3})$
result in reducible representations and have to be excluded
As above, $(k,l)$ are assumed to take their standard values
only. When summing over the classes we have similar to
Eq.~(\ref{c1sum})

\beqn
\sum_{C_1^{(\rho,\sig)}} & = & \frac{1}{3}
\left( \sum_{\rho,\sig=0}^{n-1}  ~~~-~~
\sum_{\rho=-\sig=\frac{n}{3},\frac{2n}{3}} ~~-~~~ \sum_{\rho=\sig=0}
\right).
\eeqn

Noting that the characters of $C_1 \Big/
C_1^{(\rho)}$ are identical to the characters of $C_1^{(\rho,\sig)}$
with $(\rho,\sig)=(0,0) \Big/ (\rho,\sig)=(\rho,-\rho)$,
$\rho=\frac{n}{3},\frac{2n}{3}$, respectively, we can write the sum over
the classes of type~1 in Eq.~(\ref{decomp}) as

\beqn
&&\chi^{[{\bf r}]}_{C_1} \, \chi^{[{\bf s}]}_{C_1} \,
\ol{\chi}^{[{\bf t}]}_{C_1}  ~+ \sum_{C_1^{(\rho)}}
\chi^{[{\bf r}]}_{C_1^{(\rho)}} \, \chi^{[{\bf s}]}_{C_1^{(\rho)}} \,
\ol{\chi}^{[{\bf t}]}_{C_1^{(\rho)}} ~+ \sum_{C_1^{(\rho,\sig)}} 3
\cdot \chi^{[{\bf r}]}_{C_1^{(\rho,\sig)}} \,
\chi^{[{\bf s}]}_{C_1^{(\rho,\sig)}} \,
\ol{\chi}^{[{\bf t}]}_{C_1^{(\rho,\sig)}}
\notag \\
&&~=~ \sum_{\rho,\sig=0}^{n-1}   \chi^{[{\bf r}]}_{C_1^{(\rho,\sig)}}
\, \chi^{[{\bf s}]}_{C_1^{(\rho,\sig)}} \,
\ol{\chi}^{[{\bf t}]}_{C_1^{(\rho,\sig)}}.
\eeqn

The following calculations of the products of two representations
can be further simplified by observing that

\beqn
\frac{n}{3} \cdot
M^p \begin{pmatrix} 1 \\ 1 \end{pmatrix} = \frac{n}{3} \cdot
\begin{pmatrix} 1 \\ 1 \end{pmatrix} ~\mathrm{mod}\,(n),\label{useful}
\eeqn

independent of the value of $p$.
\newpage

\begin{itemize}

\item[$\bullet$] ${\bf 1}_{r,s} \otimes {\bf 1}_{r^\prime,s^\prime}$.

With the definitions of $S\equiv s+s^\prime - s^{\prime\prime}$ and
$R\equiv r+r^\prime - r^{\prime\prime}$, we get \beqn
d({\bf 1}_{r,s},{\bf 1}_{r^\prime,s^\prime},{\bf 1}_{r^{\prime\prime},s^{\prime\prime}})
&=& \frac{1}{3n^2} \cdot \left( \sum_{\rho,\sig=0}^{n-1}
\om^{(\rho+\sig)S} + \sum_{\tau=0}^2
\frac{n^2\!\!}{3}\, \om^{\tau S} ( \om^R +  \om^{-R} )  \right) \notag \\
&=& \frac{1}{3} \cdot \delta_{S,0}^{(3)} \cdot ( 1 + \om^R +  \om^{-R} )\notag \\
&=&\delta_{S,0}^{(3)} \cdot \delta_{R,0}^{(3)} . \label{111b} \eeqn
\beqn d({\bf 1}_{r,s},{\bf 1}_{r^\prime,s^\prime},{\bf 3}_{(k,l)}) &=& \frac{1}{3n^2}
\cdot  \sum_{\rho,\sig=0}^{n-1} \sum_{p=0}^2
\om^{(\rho+\sig)(s+s^\prime)} \cdot
\eta^{-(\rho,\sig)M^p \mbox{\tiny$\begin{pmatrix} k \\l \end{pmatrix}$}}\notag\\
&=&\frac{1}{3n^2} \cdot  \sum_{\rho,\sig=0}^{n-1} \sum_{p=0}^2
\eta^{(\rho,\sig)\left[(s+s^\prime)\frac{n}{3}
\mbox{\tiny$\begin{pmatrix} 1 \\1 \end{pmatrix}$} -
M^p \mbox{\tiny$\begin{pmatrix} k \\l \end{pmatrix}$}\right]}\notag\\
&=&\frac{1}{n^2} \cdot  \sum_{\rho,\sig=0}^{n-1}
\eta^{(\rho,\sig)\left[(s+s^\prime)\frac{n}{3}
\mbox{\tiny$\begin{pmatrix} 1 \\1 \end{pmatrix}$} -
\mbox{\tiny$\begin{pmatrix} k \\l \end{pmatrix}$}\right]} ~=~
0.\label{113b} \eeqn In the second step of Eq.~(\ref{113b}) we have
absorbed $M^p$ into a redefinition of $(\rho,\sig)$ and made use of
Eq.~(\ref{useful}). As $(k,l)$ with $k=l=0,\frac{n}{3},\frac{2n}{3}$
does not label an irreducible representation, the sum over
$\rho,\sig$ yields zero. Combining the results of Eq.~(\ref{111b})
and (\ref{113b}) we find \beqn {\bf 1}_{r,s} \otimes {\bf 1}_{r^\prime,s^\prime}
& = & {\bf 1}_{r+r^\prime,s+s^\prime}. \eeqn

\item[$\bullet$] ${\bf 1}_{r,s} \otimes {\bf 3}_{(k,l) }$.

Similar to Eq.~(\ref{113b}) we get \beqn
d({\bf 1}_{r,s},{\bf 3}_{(k,l)},{\bf 1}_{r^\prime,s^\prime}) &=&\frac{1}{3n^2} \cdot
\sum_{\rho,\sig=0}^{n-1} \sum_{p=0}^2
\eta^{(\rho,\sig)\left[(s-s^\prime)\frac{n}{3}
\mbox{\tiny$\begin{pmatrix} 1 \\1 \end{pmatrix}$} + M^p
\mbox{\tiny$\begin{pmatrix} k \\l \end{pmatrix}$}\right]}~=~0 . ~~~
\label{131b} \eeqn \beqn
d({\bf 1}_{r,s},{\bf 3}_{(k,l)},{\bf 3}_{(k^\prime,l^\prime)}) &=&\frac{1}{3n^2} \cdot
\sum_{\rho,\sig=0}^{n-1} \sum_{~p,p^\prime=~0}^2 \eta^{(\rho,\sig)
\left[ s\frac{n}{3} \mbox{\tiny$\begin{pmatrix} 1 \\1
\end{pmatrix}$}
  + M^p \mbox{\tiny$\begin{pmatrix} k \\l \end{pmatrix}$}
  - M^{p^\prime} \mbox{\tiny$\begin{pmatrix} k^\prime \\l^\prime \end{pmatrix}$}
\right]} \notag \\
&=&\frac{1}{n^2} \cdot \sum_{\rho,\sig=0}^{n-1} \sum_{p=0}^2
\eta^{(\rho,\sig) \left[  M^p \left\{ \frac{sn}{3}
\mbox{\tiny$\begin{pmatrix} 1 \\1 \end{pmatrix}$}
  + \mbox{\tiny$\begin{pmatrix} k \\l \end{pmatrix}$} \right\}
  -  \mbox{\tiny$\begin{pmatrix} k^\prime \\l^\prime \end{pmatrix}$}
\right]} \notag \\
&=& \delta_{ \mbox{\tiny$\begin{pmatrix} k^\prime \\l^\prime
\end{pmatrix}$} , \widetilde{\mbox{\tiny$\begin{pmatrix} k +
{sn}/{3}   \\  l + {sn}/{3}
\end{pmatrix}$}}}.\label{133b}
\eeqn Again we have absorbed $M^{p^\prime}$ into $(\rho,\sig)$ and
applied Eq.~(\ref{useful}). As the pairs $(k^\prime,l^\prime)$ take
standard values, only one $p$ contributes to the sum.
Together, Eqs.~(\ref{131b}) and (\ref{133b}) yield \beqn {\bf 1}_{r,s}
\otimes {\bf 3}_{(k,l)} & = &
{\bf 3}_{\widetilde{\mbox{\tiny$\begin{pmatrix}k+{sn}/{3} \\
l+{sn}/{3}\end{pmatrix}$}}}. \eeqn

\item[$\bullet$] ${\bf 3}_{(k,l)} \otimes {\bf 3}_{(k^\prime,l^\prime) }$.

Up to some sign changes, the coefficient
$d({\bf 3}_{(k,l)},{\bf 3}_{(k^\prime,l^\prime)},{\bf 1}_{r,s})$  is calculated as in
Eq.~(\ref{133b}) \beqn d({\bf 3}_{(k,l)},{\bf 3}_{(k^\prime,l^\prime)},{\bf 1}_{r,s})
&=& \delta_{ \mbox{\tiny$\begin{pmatrix} k^\prime \\l^\prime
\end{pmatrix}$} , \widetilde{\mbox{\tiny$\begin{pmatrix} - k +
{sn}/{3}   \\  - l + {sn}/{3}
\end{pmatrix}$}}}.\label{331b}
\eeqn The calculation of
$d({\bf 3}_{(k,l)},{\bf 3}_{(k^\prime,l^\prime)},{\bf 3}_{(k^{\prime\prime},l^{\prime\prime})})$
is completely identical to Eq.~(\ref{333}). We therefore have
\beqn {\bf 3}_{(k,l)} \otimes {\bf 3}_{(k^\prime,l^\prime) } &=& \sum_{s=0}^2
\delta_{ \mbox{\tiny$\begin{pmatrix} k^\prime \\l^\prime
\end{pmatrix}$} , \widetilde{\mbox{\tiny$\begin{pmatrix} - k +
{sn}/{3}   \\  - l + {sn}/{3}
\end{pmatrix}$}}}  ~ ({\bf 1}_{0,s} + {\bf 1}_{1,s} + {\bf 1}_{2,s}) ~~ + \notag\\
&& +~{\bf 3}_{\widetilde{\mbox{\tiny$\begin{pmatrix} k^\prime+k \\
l^\prime+l \end{pmatrix}$}}}
+~{\bf 3}_{\widetilde{\mbox{\tiny$\begin{pmatrix} k^\prime-k-l \\
l^\prime+k  \end{pmatrix}$}}} +~
{\bf 3}_{\widetilde{\mbox{\tiny$\begin{pmatrix} k^\prime+l \\ l^\prime-k-l
\end{pmatrix}$}}}. \label{33b} \eeqn Depending on the values of
$(k,l)$ and $(k^\prime,l^\prime)$ we can have
either nine, three, or zero singlets.  \\
Let us first focus on the case where $k,l,k^\prime,l^\prime$ are all
multiples of $\frac{n}{3}$. Adopting the conventions of
Appendix~\ref{standard} for choosing the standard representatives,
$(k,l)$ and $(k^\prime,l^\prime)$ can take only two values:
$(0,\frac{n}{3})$ and $(0,\frac{2n}{3})$. For
$(k^\prime,l^\prime)=(-k,-l)$ we are left with nine singlets. If
$(k^\prime,l^\prime)=(k,l)$, we have no singlets but three
(identical)
triplets ${\bf 3}_{\widetilde{(2k,2l)}}$. \\
In all other cases, i.e. those cases where at least one of the four
parameters $k,l,k^\prime,l^\prime$ is {\it not} a multiple of
$\frac{n}{3}$, it is possible to show that the Kronecker delta can
be non-zero only for one value of $s$. Thus we obtain either three
or zero singlets. In order to have singlets, there must be a pair
$p^\prime,s=0,1,2$ such that \beqn
\begin{pmatrix} k^\prime \\ l^\prime \end{pmatrix} &=&
M^{p^\prime} \begin{pmatrix} -k \\ -l \end{pmatrix} + \frac{sn}{3}
\begin{pmatrix} 1 \\ 1 \end{pmatrix}.\label{singlets} \eeqn Then,
one of the three triplets would not exist. This can be seen by
rewriting the subscripts of the triplets
${\bf 3}_{(k^{\prime\prime},l^{\prime\prime})}$ in Eq.~(\ref{33b}) and
inserting Eq.~(\ref{singlets}): \beqn
\begin{pmatrix} k^{\prime\prime} \\  l^{\prime\prime} \end{pmatrix}
&=&  M^{p^{\prime\prime}} \left[
\begin{pmatrix} k^\prime \\ l^\prime \end{pmatrix}
+ M^p \begin{pmatrix} k \\ l \end{pmatrix}  \right] \notag \\
&=&  M^{p^{\prime\prime}} \left[ M^{p^\prime} \begin{pmatrix} -k \\
-l \end{pmatrix} + \frac{sn}{3} \begin{pmatrix} 1 \\ 1 \end{pmatrix}
+ M^p \begin{pmatrix} k \\ l \end{pmatrix}  \right] \notag \\
&=&  M^{p^{\prime\prime}} \left[
 (M^p - M^{p^\prime}) \begin{pmatrix} k \\ l \end{pmatrix} \right]
+ \frac{sn}{3} \begin{pmatrix} 1 \\ 1 \end{pmatrix} . \eeqn $p=0
\big/ 1 \big/ 2$  correspond to the first$\big/$second$\big/$third
triplet in Eq.~(\ref{33b}). The matrix $M^{p^{\prime\prime}}$ takes
care of the mapping $\widetilde{\phantom{va}}$ onto the standard
representatives. As the pairs $(k^{\prime\prime},l^{\prime\prime})=
(0,0),(\frac{n}{3},\frac{n}{3}),(\frac{2n}{3},\frac{2n}{3})$ in
Eq.~(\ref{333}) are not allowed for $n=3 \,\mathbb{Z}$, the
triplet with $p=p^\prime$ is absent and replaced by three singlets
instead.
\end{itemize}

\end{itemize}

\cleqn
\section{Clebsch-Gordan Coefficients}\label{CGapp}

Under the action of $a$, $c$, and $d$, the $\Delta(3n^2)$ triplet
fields $\vec\varphi$ and $\vec\varphi^\prime$ transform as

\beq
\begin{pmatrix} \varphi_1 \\ \varphi_2 \\ \varphi_3 \end{pmatrix}
~~ \longmapsto ~~
\begin{pmatrix} \varphi_2 \\ \varphi_3 \\ \varphi_1 \end{pmatrix}_a \ ,~~~
\begin{pmatrix} \eta^l~\varphi_1 \\ \eta^k~\varphi_2 \\ \eta^{-k-l}~\varphi_3
\end{pmatrix}_c \ ,~~~
\begin{pmatrix} \eta^{-k-l}~\varphi_1 \\ \eta^l~\varphi_2 \\ \eta^k~\varphi_3
\end{pmatrix}_d \ ,
\eeq and \beq
\begin{pmatrix} \varphi^\prime_1 \\ \varphi^\prime_2 \\ \varphi^\prime_3 \end{pmatrix}
~~ \longmapsto ~~
\begin{pmatrix} \varphi^\prime_2 \\ \varphi^\prime_3 \\ \varphi^\prime_1 \end{pmatrix}_a \ ,~~~
\begin{pmatrix} \eta^{l^\prime}~\varphi^\prime_1 \\ \eta^{k^\prime}~\varphi^\prime_2 \\ \eta^{-k^\prime-l^\prime}~\varphi^\prime_3
\end{pmatrix}_c \ ,~~~
\begin{pmatrix} \eta^{-k^\prime-l^\prime}~\varphi^\prime_1 \\ \eta^{l^\prime}~\varphi^\prime_2 \\ \eta^{k^\prime}~\varphi^\prime_3
\end{pmatrix}_d \ ,
\eeq respectively.

\begin{itemize}

\item[$(i)$] $n\neq 3\,\mathbb{Z}$.

In terms of the component fields, Eq.~(\ref{2singlets}) is written
as \beq \varphi^\prime_1 \varphi_1 + \om^{-r} \, \varphi^\prime_2
\varphi_2 + \om^{r} \, \varphi^\prime_3 \varphi_3 \ .\label{two1s}
\eeq For $(k^\prime,l^\prime)=(-k,-l)$ this expression is invariant
under $c$ and $d$. Under the action of $a$, Eq.~(\ref{two1s}) is
transformed to \beq \varphi^\prime_2 \varphi_2 + \om^{-r} \,
\varphi^\prime_3 \varphi_3 + \om^{r} \, \varphi^\prime_1 \varphi_1
~=~ \om^{r} \, ( \, \varphi^\prime_1 \varphi_1 + \om^{-r} \,
\varphi^\prime_2 \varphi_2 + \om^{r} \, \varphi^\prime_3 \varphi_3
\,) \ .\label{cgderivation} \eeq The Clebsch-Gordan coefficients are
thus given as \beqn \langle {\bf 3}_{(k^\prime,l^\prime)}^{i^\prime}
\,,\, {\bf 3}_{(k,l)}^i \,|\, {\bf 1}_r \rangle &=&
\om^{r(1-i)}\;\delta_{i^\prime,i} \;
\delta_{\mbox{\tiny$\begin{pmatrix}
k^\prime\\l^\prime\end{pmatrix}$}, \mbox{\tiny$\begin{pmatrix} -k \\
-l \end{pmatrix}$}}\ , \eeqn with $i=1,2,3$ denoting the component
of the triplet $\vec\varphi$.

Turning to the three-dimensional representations build out of the
product of two triplets, let us consider the three vectors \beq
\begin{pmatrix} \varphi^\prime_1 \varphi_1 \\  \varphi^\prime_2 \varphi_2 \\
  \varphi^\prime_3 \varphi_3   \end{pmatrix} \ , ~~
\begin{pmatrix} \varphi^\prime_1 \varphi_2 \\  \varphi^\prime_2 \varphi_3 \\
  \varphi^\prime_3 \varphi_1   \end{pmatrix} \ , ~~
\begin{pmatrix} \varphi^\prime_1 \varphi_3 \\  \varphi^\prime_2 \varphi_1 \\
  \varphi^\prime_3 \varphi_2   \end{pmatrix}\ ,\label{threedimvec}
\eeq which under $c$ transform as \beq
\begin{pmatrix} \eta^{l^\prime+l}~\varphi^\prime_1 \varphi_1 \\
  \eta^{k^\prime+k}~\varphi^\prime_2 \varphi_2 \\
\eta^{-k^\prime-l^\prime-k-l}~\varphi^\prime_3 \varphi_3
\end{pmatrix} \ , ~~
\begin{pmatrix} \eta^{l^\prime+k}~\varphi^\prime_1 \varphi_2 \\
  \eta^{k^\prime-k-l}~\varphi^\prime_2 \varphi_3 \\
  \eta^{-k^\prime-l^\prime+l}~\varphi^\prime_3 \varphi_1   \end{pmatrix} \ , ~~
\begin{pmatrix} \eta^{l^\prime-k-l}~\varphi^\prime_1 \varphi_3 \\
  \eta^{k^\prime+l}~\varphi^\prime_2 \varphi_1 \\
  \eta^{-k^\prime-l^\prime+k}~\varphi^\prime_3 \varphi_2   \end{pmatrix}\ .
\label{ctrafo3dimvec} \eeq Under the action of $a$, the components
of the three-dimensional vectors in Eq.~(\ref{threedimvec}) are
permuted cyclically. Hence, up to cyclic permutations, these vectors
constitute the triplet representations ${\bf
3}_{(k^{\prime\prime},l^{\prime\prime})}$ of $\Delta(3n^2)$ with
$(k^{\prime\prime},l^{\prime\prime})$ defined by
Eq.~(\ref{ctrafo3dimvec}). The ambiguity of labeling the
three-dimensional representations is taken care of by introducing
the parameter $p$; as in Eq.~(\ref{kpplppGEN}) we can write \beqn
\begin{pmatrix} k^{\prime\prime} \\ l^{\prime\prime} \end{pmatrix} &=& M^p
\left[ \begin{pmatrix} k^\prime \\ l^\prime \end{pmatrix} +
\begin{pmatrix} k \\ l \end{pmatrix} \right]\ , \label{appendixkpplpp}
\eeqn for the first of the three vectors in Eq.~(\ref{threedimvec}).
The corresponding three-dimensional representation reads either \beq
\begin{pmatrix} \varphi^\prime_1 \varphi_1 \\  \varphi^\prime_2 \varphi_2 \\
  \varphi^\prime_3 \varphi_3   \end{pmatrix}_{p=0} \ , ~~~\mathrm{or}~~~
\begin{pmatrix}   \varphi^\prime_2 \varphi_2 \\ \varphi^\prime_3 \varphi_3 \\
  \varphi^\prime_1 \varphi_1  \end{pmatrix}_{p=1} \ , ~~~\mathrm{or}~~~
\begin{pmatrix}   \varphi^\prime_3 \varphi_3  \\ \varphi^\prime_1 \varphi_1 \\
  \varphi^\prime_2 \varphi_2  \end{pmatrix}_{p=2} \ .
\eeq Therefore, we get the Clebsch-Gordan coefficient as \beqn
\langle {\bf 3}_{(k^\prime,l^\prime)}^{i^\prime} \,,\, {\bf
3}_{(k,l)}^i \,|\, {\bf
3}_{(k^{\prime\prime},l^{\prime\prime})}^{i^{\prime\prime}} \rangle
&=& \delta^{(3)}_{i^{\prime\prime}, i^\prime-p}
\;\delta_{i^\prime,i} \; \delta_{\mbox{\tiny$\begin{pmatrix}
 k^{\prime\prime}\\l^{\prime\prime}\end{pmatrix}$},
{\mbox{\tiny$\,M^p\begin{pmatrix} k^\prime +k
 \\ l^\prime + l \end{pmatrix}$}}}\ ,
\eeqn with $p$ defined in Eq.~(\ref{appendixkpplpp}). The
superscript $^{(3)}$ on the Kronecker delta indicates that we
calculate modulo~3. Generalizing this result to the other two
vectors of Eq.~(\ref{threedimvec}) we find \beqn \langle {\bf
3}_{(k^\prime,l^\prime)}^{i^\prime} \,,\, {\bf 3}_{(k,l)}^i \,|\,
{\bf 3}_{(k^{\prime\prime},l^{\prime\prime})}^{i^{\prime\prime}}
\rangle &=&
\delta^{(3)}_{i^{\prime\prime},i^\prime-p}\;\delta^{(3)}_{i^\prime,i-q}
\; \delta_{\mbox{\tiny$\begin{pmatrix}
 k^{\prime\prime}\\l^{\prime\prime}\end{pmatrix}$},
 {\mbox{\tiny$\,M^p \left[\begin{pmatrix} k^\prime \\ l^\prime \end{pmatrix} +
 M^q \begin{pmatrix} k \\ l \end{pmatrix} \right]$}}}\ .\label{cgtrip}
\eeqn If $(k^\prime,l^\prime)\neq(-k,-l)$, one can build three
($q=0,1,2$) different three-dimensional representations  out of the
product of two  triplets. In the case where
$(k^\prime,l^\prime)=(-k,-l)$, only $q=1,2$ yield triplet
representations; the remaining degrees of freedom are the three
singlet states of Eq.~(\ref{two1s}).

\item[$(ii)$] $n = 3\,\mathbb{Z}$.

The one-dimensional representations obtained from the product of two
triplets are given in Eqs.~(\ref{2singletsa}) and
(\ref{2singletsb}). In terms of the component fields, they take the
form \beqn &\varphi^\prime_1 \varphi_1 + \om^{-r} \,
\varphi^\prime_2 \varphi_2 +
\om^{r} \, \varphi^\prime_3 \varphi_3 \ ,\label{cgsing1} \\[3mm]
&\varphi^\prime_2 \varphi_1 + \om^{-r} \, \varphi^\prime_3 \varphi_2
+
\om^{r} \, \varphi^\prime_1 \varphi_3 \ ,\label{cgsing2} \\[3mm]
&\varphi^\prime_3 \varphi_1 + \om^{-r} \, \varphi^\prime_1 \varphi_2
+ \om^{r} \, \varphi^\prime_2 \varphi_3 \ .\label{cgsing3} \eeqn
Analogous to Eq.~(\ref{cgderivation}), the action of $a$ changes
these expression only by an overall factor of $\om^r$. With regard
to the action of the generator $c$, we have to specify the values of
$k^\prime,l^\prime,k,l$. We treat the two cases of
Section~\ref{invariants} separately.

\begin{itemize}
\item[$(a)$] In this case [cf. Eq.~(\ref{2singletsa})] we have
  $(-k^\prime,-l^\prime)=(k,l)=(0,\pm n/3)$. Recalling that $\eta^{n/3} =
  \om$ and $\om^3=1$, Eqs.~(\ref{cgsing1})-(\ref{cgsing3}), under the action of
  $c$, transform as
\beqn &\varphi^\prime_1 \varphi_1 + \om^{-r} \, \varphi^\prime_2
\varphi_2
+  \om^{r} \, \varphi^\prime_3 \varphi_3 \ , \\[3mm]
&\om^{\pm 1} \, (\varphi^\prime_2 \varphi_1 + \om^{-r} \,
\varphi^\prime_3 \varphi_2 +
\om^{r} \, \varphi^\prime_1 \varphi_3) \ ,\\[3mm]
&\om^{\mp 1} \, (\varphi^\prime_3 \varphi_1 + \om^{-r} \,
\varphi^\prime_1 \varphi_2 + \om^{r} \, \varphi^\prime_2 \varphi_3)
\ , \eeqn where the two signs in the exponents correspond to the two
possible values for $l=\pm n/3$. Hence, the expression in
Eq.~(\ref{cgsing1}) gives the ${\bf 1}_{r,0}$ representation. For
$l=+n/3$, the  ${\bf 1}_{r,1}$ representation is given by
Eq.~(\ref{cgsing2}) and the ${\bf 1}_{r,2}$ representation by
Eq.~(\ref{cgsing3}). This is opposite for the case where $l=-n/3$.
The Clebsch-Gordan coefficients are thus given as \beqn \langle {\bf
3}_{(k^\prime,l^\prime)}^{i^\prime} \,,\, {\bf 3}_{(k,l)}^i \,|\,
{\bf 1}_{r,s} \rangle &=& \om^{r(1-i)}\;\delta^{(3)}_{i^\prime,i \pm
s} \; \delta_{\mbox{\tiny$\begin{pmatrix}
k^\prime\\l^\prime\end{pmatrix}$}, \mbox{\tiny$\begin{pmatrix} -k \\
-l \end{pmatrix}$}}\ , \eeqn with $(k,l)=(0,\pm n/3)$.

\item[$(b)$] Here, $(k^\prime,l^\prime)$ are given in Eq.~(\ref{kplps}); they
  depend on $s$ and the mapping $\widetilde{\phantom{va}}$
  (i.e. the value of $p$). For $p=0$, Eq.~(\ref{cgsing1}) transforms as
\beq \om^s \, (\varphi^\prime_1 \varphi_1 + \om^{-r} \,
\varphi^\prime_2 \varphi_2 +  \om^{r} \, \varphi^\prime_3 \varphi_3)
\ , \eeq under the action of $c$. It therefore constitutes the
one-dimensional representation  ${\bf 1}_{r,s}$. Similarly, for
$p=1/2$, the ${\bf 1}_{r,s}$ representation is given by
Eq.~(\ref{cgsing3})/(\ref{cgsing2}), respectively. Thus the
Clebsch-Gordan coefficients are \beqn \langle {\bf
3}_{(k^\prime,l^\prime)}^{i^\prime} \,,\, {\bf 3}_{(k,l)}^i \,|\,
{\bf 1}_{r,s} \rangle &=& \om^{r(1-i)}\;\delta^{(3)}_{i^\prime,i -p}
\; \delta_{\mbox{\tiny$\begin{pmatrix}
k^\prime\\l^\prime\end{pmatrix}$}, \mbox{\tiny$M^p \begin{pmatrix}
-k +sn/3\\ -l +sn/3 \end{pmatrix}$}}\ . \eeqn
\end{itemize}

The three-dimensional representations are constructed as in case
($i$). Eq.~(\ref{cgtrip}) shows the corresponding Clebsch-Gordan
coefficients. In subcase ($a$), the parameter $p$ can be directly
related to $q$. This can be seen as follows: In order to obtain
triplet representations, we must have $(k^\prime,l^\prime)=(k,l)$ as
discussed below Eq.~(\ref{33b}). Taking $(0,\pm\frac{n}{3})$ as the
standard labeling for the three-dimensional representations, we can
determine $p$ for a given value of $q$ explicitly \beqn
\begin{pmatrix} k^{\prime\prime} \\ l^{\prime\prime} \end{pmatrix}
~=~ M^p \left[ \begin{pmatrix} k^\prime \\ l^\prime \end{pmatrix} +
M^q
  \begin{pmatrix} k \\ l \end{pmatrix} \right]
&=& M^p \left[ \begin{pmatrix} 0 \\ l \end{pmatrix} + M^q
  \begin{pmatrix} 0 \\ l \end{pmatrix} \right].
\eeqn $p$ is obtained by the condition that $k^{\prime\prime}=0$.
This yields the relation $p = q $, for subcase ($a$).

\end{itemize}





\begin{thebibliography}{99}

\bibitem{GLM} F. G\"ursey, and G. Feinberg, Phys.\ Rev.\, {\bf 128},
378 (1962);
  T.~D.~Lee, Nuovo Cim.\  {\bf 35}, 975 (1965); S.~Meshkov and S.~P.~Rosen,
  Phys.\ Rev.\ Lett.\  {\bf 29}, 1764 (1972).

\bibitem{WZ}
  F.~Wilczek and A.~Zee,
  Phys.\ Rev.\ Lett.\  {\bf 42}, 421 (1979).


\bibitem{CF} T.~L.~Curtright and P.~G.~O.~Freund, in
Supergravity, ed. P. van Nieuwenhuizen and D. Z. Freedman, (North
Holland, Amsterdam), 1979; M. Gell-Mann, P. Ramond and R. Slansky,
{\em ibid} ; M. Bowick, and P. Ramond, Phys.\ Lett.\ B{\bf 103}:338,
1981.

\bibitem{Pakvasa}
  S.~Pakvasa and H.~Sugawara,
  Phys.\ Lett.\ B {\bf 73}, 61 (1978).

\bibitem{Schmaltz} D.~B.~Kaplan and M.~Schmaltz, Phys.\ Rev.\ D {\bf 49}, 3741
(1994); M.~Schmaltz, Phys.\ Rev.\ D {\bf 52}, 1643 (1995).

\bibitem{Ma}E.~Ma and G.~Rajasekaran,
Phys.\ Rev.\ D {\bf 64}, 113012 (2001); K.~S.~Babu, E.~Ma and
J.~W.~F.~Valle, Phys.\ Lett.\ B {\bf 552}, 207 (2003); A.~Zee,
  Phys.\ Lett.\ B {\bf 630}, 58 (2005); G.~Altarelli and F.~Feruglio,
  Nucl.\ Phys.\ B {\bf 720}, 64 (2005); G.~Altarelli and F.~Feruglio,
  Nucl.\ Phys.\ B {\bf 741}, 215 (2006); E.~Ma,
Mod.\ Phys.\ Lett.\ A {\bf 21}, 1917 (2006).


\bibitem{MNY}
W.~Grimus and L.~Lavoura, JHEP {\bf 0601}, 018 (2006);  R.~Jora,
S.~Nasri and J.~Schechter, Int.\ J.\ Mod.\ Phys.\ A {\bf 21}, 5875
(2006); R.~N.~Mohapatra, S.~Nasri and H.~B.~Yu, Phys.\ Lett.\ B {\bf
639}, 318 (2006).

\bibitem{Ross}
I.~de Medeiros Varzielas, S.~F.~King and G.~G.~Ross,
  Phys.\ Lett.\ B {\bf 644}, 153 (2007); I.~de Medeiros Varzielas, S.~F.~King
  and G.~G.~Ross, Phys.\ Lett.\ B {\bf 648}, 201 (2007).

\bibitem{Kobayashi}
   G.~Altarelli, F.~Feruglio and Y.~Lin,
arXiv:hep-ph/0610165; T.~Kobayashi, H.~P.~Nilles, F.~Pl\"oger,
S.~Raby and M.~Ratz, Nucl.\ Phys.\  B {\bf 768}, 135 (2007).

\bibitem{Miller} G. A. Miller , H. F. Dickson, and L. E. Blichfeldt,
Theory and Applications of Finite Groups, John Wiley \& Sons, New
York, 1916, and Dover Edition, 1961 .

\bibitem{Fairbairn:1964}
W.~M. Fairbairn, T.~Fulton, and W.~H. Klink, J.\ Math.\ Phys.\, {\bf
5}:1038, 1964.

\bibitem{Bovier:1980gc}
A.~Bovier, M.~L{\"u}ling, and D.~Wyler, J.\ Math.\ Phys.\, {\bf
22}:1543, 1981.






\end{thebibliography}

\end{document}